\documentclass[preprint2]{proto}
\usepackage{times}

\usepackage{natbib}          

\begin{document}

\title{\textbf{\LARGE Characterizing Exoplanet Habitability}}

\author{\textbf{\large Ravi kumar Kopparapu}}
\affil{\small\em NASA Goddard Space Flight Center}
\author{\textbf{\large Eric T. Wolf}}
\affil{\small\em University of Colorado, Boulder}
\author{\textbf{\large Victoria S. Meadows}}
\affil{\small\em University of Washington}

\begin{abstract}
\begin{list}{ } {\rightmargin 1in}
\baselineskip = 11pt
\parindent=1pc

{\small 
Habitability is a measure of an environment's potential to support life, and a habitable exoplanet supports liquid water on its surface.  However, a planet's success in maintaining liquid water on its surface is the end result of a complex set of interactions between planetary, stellar, planetary system and even Galactic characteristics and processes, operating over the planet's lifetime.  In this chapter, we describe how we can now determine which exoplanets are most likely to be terrestrial, and the research needed to help define the habitable zone under different assumptions and planetary conditions. We then move beyond the habitable zone concept to explore a new framework that looks at far more characteristics and processes, and provide a comprehensive survey of their impacts on a planet's ability to acquire and maintain habitability over time. We are now entering an exciting era of terrestiral exoplanet atmospheric characterization, where initial observations to characterize planetary composition and constrain atmospheres is already underway, with more powerful observing capabilities planned for the near and far future. Understanding the processes that affect the habitability of a planet will guide us in discovering habitable, and potentially inhabited, planets. 
\\~\\~\\~}


\end{list}
\end{abstract}

\textit{There are countless suns and countless earths all rotating around their suns in exactly the same way as the seven planets of our system. We see only the suns because they are the largest bodies and are luminous, but their planets remain invisible to us because they are smaller and non-luminous. The countless worlds in the universe are no worse and no less inhabited than our earth.}

---Giordano Bruno, 1584 A.D.

\section{\textbf{INTRODUCTION}}

 Statistical studies of the thousands of known exoplanets suggest that the majority of stars host planetary systems \citep{Cassan2012, Dressing2015, Gaidos2016, Winn2018}, and so it seems inconceivable that the Earth is the only habitable world in the Universe, even though that may indeed be true.  One of the primary goals of both exoplanet science and astrobiology is to search for and identify a potentially habitable, and possibly inhabited planet orbiting another star. For an exoplanet, habitability is defined as the ability to support and maintain liquid water on the planetary surface.  There are several extrasolar planets that are currently considered to be prime candidates for follow-up observations to determine their habitability potential, and future discoveries may yield even more candidate habitable worlds, increasing the odds of finding life outside our Solar system.  New NASA mission concepts currently under consideration are designed to have the capability to characterize the most promising planets for signs of habitability and life.   We are at an exhilarating point in human history where the answer to the question ``Are we alone?'' lies within our scientific and technological grasp. 


To understand habitability more broadly for exoplanets, however, we need to better understand how stars both like and unlike the Sun impact planetary environments. The habitability potential of a planet critically depends upon the host star’s characteristics, which can include: stellar spectral energy distribution, activity, stellar winds, age, X-ray/UV emission, magnetic field, and stellar multiplicity. Several of these factors may also change with the age of the star, consequently affecting habitability of a planet over time.  Many of these factors become  particularly critical for M dwarf habitable zone planets, which orbit much closer to their parent stars than the Earth does to the Sun.  

In addition to host star properties, habitability of a planet is also influenced by the properties and processes of the planet itself, which include but are not limited to  atmospheric composition, atmospheric escape/retention, volatile inventory and delivery, cycling of elements between surface and interior, planetary magnetic field, planet mass and size, orbital architecture of planets in the system, and the presence of giant planets. Life itself may also have an influence on the habitability of a planet \citep[see also Chapter 4 by St\"ueken et al in this volume]{Nisbet2007}. Within our Solar system there is a diversity of planetary environmental conditions, with Earth as the only known planet with surface liquid water. Our closest neighbors, Mars and Venus, seem to have taken different evolutionary paths than the Earth, primarily in response to the influence of our changing Sun over the last 4.5 billion years, but also due to geological factors.  There is evidence that Mars had flowing water on its surface 3.5 Gyr ago \citep{Fassett2008}, and it is hypothesized that Venus may have had liquid water, however the evidence remains unclear \citep{Donahue1982,Grinspoon1993,Kulikov2006,Hamano2013,Way2016}. Nevertheless, the implication that the Solar system may have had at least two planets with liquid water on their surface (and so perhaps potentially habitable) at some point in ancient times, raises an interesting possibility of similar history on planets around other stars. 

In this chapter, we will address some of the requirements for understanding and assessing planetary habitability, emphasizing that this assessment is specifically focused on exoplanets. Consequently, the surveys and measurements needed to explore the habitability potential of a planet are quite different than solar system planets, which are discussed in earlier chapters. Without a great technological leap forward, we cannot send satellites and landers to study exoplanets at close range, as has been done for virtually all of the major objects in our Solar System.  All knowledge of habitable exoplanets must be obtained via astronomical observations, and our understanding of the planetary and stellar factors that control habitability must be used to interpret these data. 

\section{IDENTIFYING POTENTIALLY HABITABLE EXOPLANETS}

\begin{figure*}
  \centering
\includegraphics[scale=0.35]{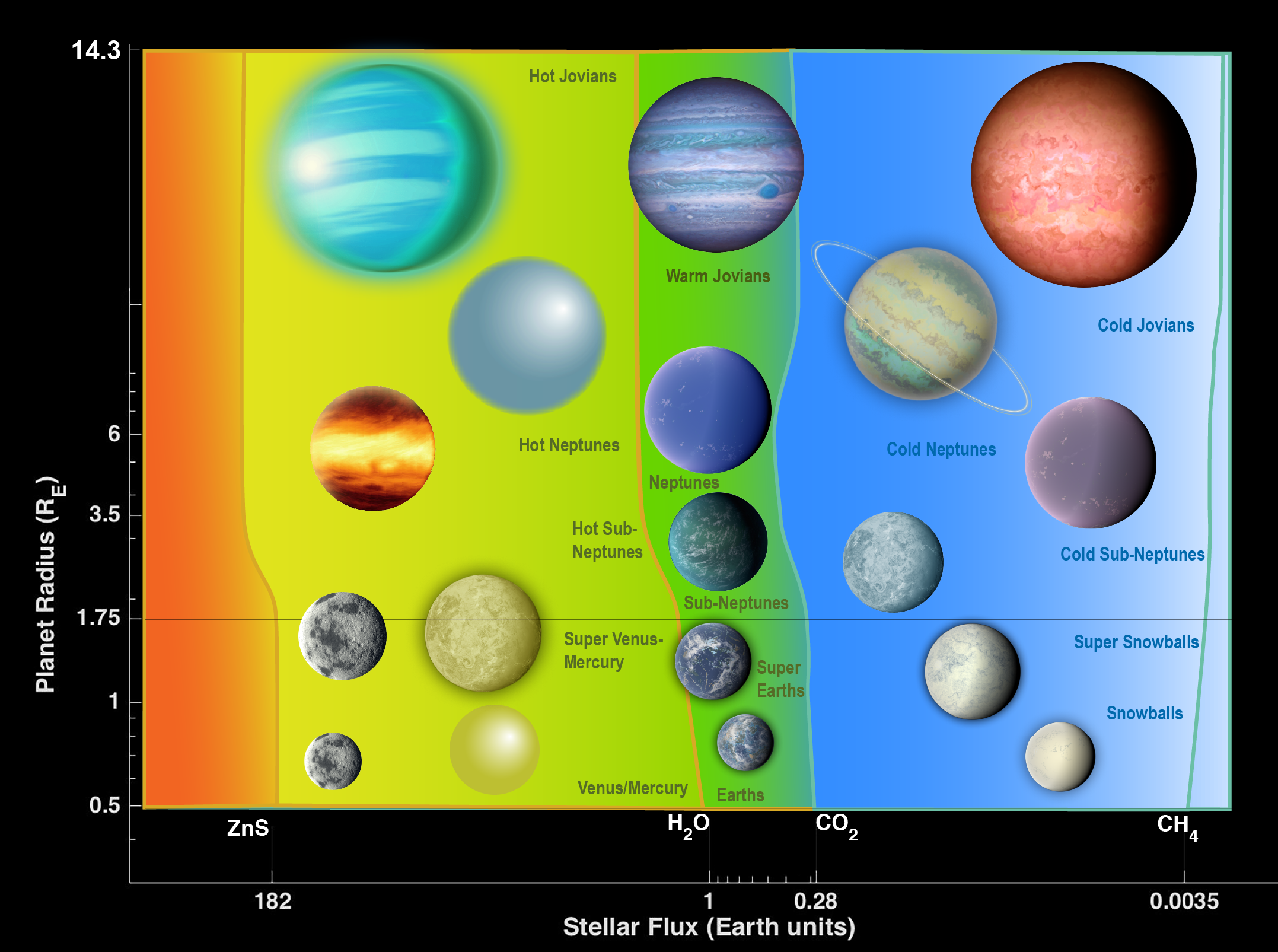}
\caption{\small{\textit{ Classification of exoplanets into different categories \citep{Kopparapu2018}. The boundaries of the boxes represent the regions where different chemical species are condensing in the atmosphere of that particular sized planet at that stellar flux, according to equilibrium chemistry calculations. The radius division is from \citet{Fulton2017} for super-Earths and sub-Neptunes, and from \citet{ChenKipping2017} for the upper limit on Jovians..}}}
\label{fig:1} 
\end{figure*}

For exoplanets, a habitable planet is defined as one that can support liquid water on its surface.  This ``surface liquid water'' criterion has been used to define the Habitable Zone (HZ; \citet{Hart1978,Hart1979}) as that range of distances from a parent star in which an Earth-like planet could maintain liquid water on its surface \citep{Kasting1993,Kopparapu2013a} and so potentially host a surface biosphere. Although subsurface liquid water is entirely possible and may even by common---as suggested by the interior oceans of the icy moons in our Solar System---detecting that water, and any subsurface biosphere supported by it, is far less likely with remote-sensing telescopic observations (for more detailed discussions of the habitability potential of the surface and subsurface environments of Mars and the Solar System's icy moons see Chapters 6-9 by Amador et al., Davila et al., Schmidt and Cable et al., in this volume).  Consequently, the search for habitability and life on exoplanets will focus on telescopic observations of planetary atmospheres and surfaces, where a surface biosphere will be more apparent.  

The habitable zone is therefore designed as a useful concept to identify that region around a star where an orbiting planet has the highest probability of being \textit{detectably} habitable, for remote-sensing studies.  Although we do not currently have a means of observing markers of surface habitability on exoplanets, these capabilities are expected in the near future (see Section \ref{sec:observations}).  Arguments that the habitable zone is somehow too limited, because it does not encompass the subsurface habitability exemplified by the Solar System's icy moons (e.g. \citet{Stevenson2018}, \citet{Tasker2017}), do not take into account the definition and purpose of the habitable zone. 

In the search for habitable exoplanets, it is an important first step to be able to identify those planets that are most likely to be habitable.  These planets will become the highest priority targets for future telescopes that will be able to observationally confirm whether or not a planet supports liquid surface water.  A first order assessment of potential habitability would be to 1) find a planet that has the solid surface needed to support an ocean, and 2) that resides within the habitable zone, so that liquid surface water is more likely to be possible.  This initial assessment can be made with three readily observable characteristics: the planet's mass or size, the type of star it orbits, and its distance from that star. 

However, as the field of astrobiology develops, it is becoming clearer that multiple factors, characteristics and processes, can impact whether a planet is able to acquire and maintain liquid water on its surface.  These include the properties of the planet, star and planetary system, and how these interact over time \citep{MeadowsBarnes2018}. Finding a terrestrial-type rocky planet in the habitable zone can then be thought of as  a two-dimensional slice through a far more complex, interdisciplinary and multi-dimensional parameter space. Moreover, a planet's position in the HZ does not guarantee habitability, because aspects of its formation or evolution may preclude habitability.  For example, the planet could have formed with little or no water \citep{Raymond2004,Raymond2007}, or lost that water in the first billion years of the star's evolution \citep{RamirezKalt2014, Luger2015a, Tian&Ida2015}. In the rest of this section we discuss the larger context of types of exoplanets, and how we now feel confident we can identify those planets most likely to be terrestrial, and also expand our discussion of how the habitable zone is defined.  In subsequent sections we review the many factors that can impact exoplanet habitability more broadly, and conclude with a discussion of future work in this area. 

 
\subsection{The Search for Terrestrial Exoplanets}

Exoplanet discoveries have revealed a diversity of exoplanets that span a broad range of mass/radius and orbital distance, exceeding the types of planets seen in our Solar System, and  arrayed in planetary system architectures that are often completely unlike our own \citep{Fulton2017, Winn2018}.  While direct analogs of gas giants, ice giants and terrestrial planets likely exist in other systems, the exoplanet population has also revealed hot Jupiters, Jovian planets in extremely short orbital periods ($\sim$few hours to days); hot Earths, planets that are likely rocky, and that receive many more times the insolation received by Mercury (e.g. \citet{Berta2015}); and warm Neptunes that have ice giant sizes and densities, but reside within the inner planetary system.  Perhaps the most unexpected discovery has been that of the sub-Neptune population of exoplanets.  These planets are smaller than Neptune and often larger than Earth, and are currently seen in relatively close orbits ($\sim$ 200 days or less). The sub-Neptunes are a type of planet that has no analog in our Solar System, and they are extremely common, currently comprising the largest fraction of the known population of exoplanets. These sub-Neptunes appear to consist of two sub-groups divided by composition, and potentially formation mechanisms:  mini-Neptunes that are ice-dominated, and super-Earths that have densities more consistent with rock, and so may well be terrestrial exoplanets \citep{Rogers2015,Fulton2017}. 

To date, the vast majority of exoplanets have been discovered by indirect detection, that is, the presence of the planet is inferred from the behaviour of the star---which may dim, brighten or move under the influence of its orbiting planet.  For a review of the four principal indirect detection techniques, including radial velocity, transit, astrometry and microlensing, see  \citet{Fischer2014Detection,Wright2018}).  Here we will only further discuss the two current principal indirect detection techniques---radial velocity and transit---as well as direct detection techniques that isolate photons from the planet itself, such as direct imaging and secondary eclipse.  

Radial velocity was the first exoplanet detection technique to discover multiple planets orbiting main sequence stars.  The radial velocity technique detects the presence of the planet when the planet and star orbit a mutual center of mass, which causes the star to appear to move towards and away from the observer  To detect the star's radial motion (towards and away), high-resolution spectroscopy is used with ultrastable spectral reference frames, e.g. iodine gas cells, to detect the star's tiny shifts in radial velocity.  Because the radial velocity amplitude is proportional to the mass of the planet and inversely proportional to the planet's orbital distance and the mass of the star, the RV technique is particularly sensitive to large planets close to small parent stars. RV measurements can reveal planetary orbital period, eccentricity, and put constraints on planetary mass, deriving a minimum mass---which is uncertain due to the often unknown orbital inclination of the planet with respect to the observer.  Radial velocity was initially the most successful planet detection technique, and in 1995 it was used to detect the first exoplanet around a main sequence star, 51 Pegasus b \citep{Mayor1995}. Given the detection sensitivity biases of this technique, it should be no surprise that 51 Peg b was a hot Jupiter, a large planet close to its parent star.  Many other hot Jupiters were initially discovered by the RV technique, again due to the sensitivity bias, even though we now know that systems housing hot Jupiters are rare, comprising only approximately 1\% of planetary systems \citep{Wright2012frequency}.  The RV technique has continued to push down to smaller masses, and Earth-sized planets were eventually discovered orbiting M dwarfs using this technique, perhaps the most notable being Proxima Centauri b, a 1.3 M$_\oplus$ minimum mass planet orbiting the nearest star to our Sun \citep{Anglada-Escude2016}.  Radial velocity currently lacks the sensitivity to be able to detect an Earth-like planet orbiting a Sun-like star, but new community initiatives in Extreme Precision Radial Velocity (EPRV) will tackle this challenge in the coming decade. 

The launch of the \textit{Kepler} spacecraft in 2009 ushered in the heydey of planet detection using the transit technique \citep{Borucki2010}.  Unlike the RV technique which looks for tiny motions of the star, the transit technique detects planets via observation of periodic dimming of the parent star as a planet passes in front of it along our line of sight. Like the RV technique, transit is most sensitive to larger planets (that can block more light) orbiting smaller stars (so that a larger percentage of light is blocked) and it also favors planets that are closer to the parent star.  The latter attribute is valuable both because closer planets have a higher probability of appearing to transit their star relative to the observer, and because closer planets have shorter orbital periods and so multiple transits, which make for a more robust detection, occur in shorter intervals of time. The transit technique can measure the orbital period and size of the planet.  If the transiting planet can also be detected with RV, then the orbital inclination of the planet with respect to the observer, as determined by the transit geometry, removes the ambiguity on the RV mass, allowing a true mass to be inferred.  With mass from RV and size from transit, density can be calculated for the planet, which can be used to help constrain planetary bulk composition.  In multi-planet transiting systems, gravitational interactions between planets can delay or accelerate the time of subsequent transits, resulting in Transit Timing Variations (TTVs) that can also be used to infer planetary mass \citep{Agol2005, Holman2005, Winn2010}. The first detection of a transiting planet was made in 1999, the hot Jupiter HD209458 b \citep{Charbonneau2000}. In the subsequent decade, ground-based telescopes continued to make tens of transiting planet discoveries, detecting planets ranging in size from hot Jupiters to the mini-Neptune GJ1214b \citep{Charbonneau2009}.  The launch of the dedicated transit-detection space telescope, \textit{Kepler} in 2009, has pushed exoplanet detection into the thousands, rapidly eclipsing the planets detected by the RV technique.  The transit technique can also find terrestrial-sized planets orbiting in the habitable zone of their parent M dwarfs, with the TRAPPIST-1 system of seven Earth-like planets orbiting a late type (the smaller and cooler end of the spectral class) M dwarf as a key example \citep{Gillon2016,Gillon2017,Luger2017c}. However, transit also finds detecting Earth-like planets orbiting more Sun-like stars more challenging, although the upcoming PLATO mission has a main objective to determinine the bulk properties and ages of small planets, including those in the habitable zone of Sun-like stars \citep{Rauer2014} 

Direct detection provides another suite of techniques that can be used to both detect and characterize exoplanets, and that can study planets beyond the inner planetary regions favored by RV and transit measurements.  In direct detection, photons from the planet are separated from the star either spatially, with direct imaging, or temporally, using secondary eclipse, where a planet passes behind its parent star. In direct imaging, telescopes of sufficient size have the ability to angularly separate the planet from the star on the sky, and some form of starlight suppression technique is used to reduce the glare from the parent star so that the planet and star can be seen as two separate points of light.  This allows studies of the planet using both direct reflected light photometry and spectroscopy of the planet's atmosphere and surface, if it has one. To date, direct imaging has been successful only for tens of young (and so still hot and self-luminous) Jovian planets in the outer regions of planetary systems \citep{Marois2008, Marois2010, Rajan2017}. Future observations of Neptune-sized objects closer to the star may be possible with coronagraphs on board {\it James Webb Space telescope} (JWST) \citep{Beichman2019}.  Direct imaging of terrestrial planets in the habitable zone of M dwarfs may be possible for a handful of the nearest M dwarf planets in the near term \citep{Quanz2015,Crossfield2016, Lopez-Morales2019}.  On longer timescales, the imaging of terrestrial planets in the habitable zones of more Sun-like stars, and of the cool Jovians that characterize our own planetary system  will require large aperture space-based telescopes like the HabEx and LUVOIR  mission concepts\footnote{\url{https://www.greatobservatories.org/}}.   

Secondary eclipse is a means of separating a transiting planet's emitted photons from the parent star's, without planet and star needing to be spatially resolved.  The secondary eclipse technique uses observations of the unresolved star and planet, and then subtracts an observation of the star alone, taken when the planet is behind the star.  This isolates radiation from the planet, and this technique is most effective at mid-infrared wavelengths where the emitted contrast between planet and star is relatively larger, and thus easier to differentiate.  Because emitted radiation is being measured, secondary eclipse is sensitive to planetary temperature, and emission spectroscopy can also be used to measure atmospheric molecules. 

The statistics provided by the many exoplanet detection techniques, and by \textit{Kepler} in particular, have enabled many seminal exoplanet discoveries. The majority of planets detected by \textit{Kepler} are larger than our Earth, but smaller than Neptune, and reside close-in to their host stars \citep{Winn2018}). Note that this does not necessarily mean that sub-Neptunes larger than Earth are the most common type of planet in the Galaxy, as the \textit{Kepler} survey is not sensitive enough to detect smaller terrestrial planets that are potentially more numerous.  We have also learned that many planetary systems are not like our Solar System, either because they contain hot Jupiters or sub-Neptunes close to the star, or because they are systems where multiple planets are packed much closer to the star than Mercury is to the Sun \citep{Lissauer2011}.  

In the past few years, astronomers have made significant progress in understanding the nature of the sub-Neptunes, and most importantly for astrobiology, in identifying  likely terrestrial planets in this population.  Using the sample of small \textit{Kepler} planets that also have RV measurements, such that the mass, radius and density were known, researchers have applied  Bayesian statistics to help identify the dividing line in radius that corresponds to a higher likelihood that a planet smaller than that radius has a rocky composition, whereas one larger is likely to be dominated by ice or gas  \citep{Weiss2014, Rogers2015,Wolfgang2016, ChenKipping2017}. The radius below which an exoplanet is more likely to be composed of rock and metal, and so could support a surface ocean, is between 1.5-1.7  Earth radii \citep{Rogers2015}. This upper limit for terrestrial planet size was supported by more precise ground-based measurements of stellar, and therefore planetary, radii for over 1000 Kepler planet-hosting stars.  This more precise dataset found a gap in the radius distribution that had previously been washed out by the larger errors on planetary radii, that now divided sub-Neptune planets into two populations with R $< 1.5$ R$_\oplus$ and R=2.0-3.0 R$_\oplus$,   \citep{Fulton2017}.  

These two populations are inferred to be terrestrial, rocky planets, and a cohort of larger planets that have rocky cores augmented by significant gas envelopes.  This inference is due in part to the previous research that showed that planets smaller than 1.5 R$_\oplus$ had densities consistent with terrestrial planets, but also because a gap in the radius distribution of sub-Neptune planets was predicted as a result of photoevaporation of planetary envelopes by X-ray and extreme ultraviolet (XUV) radiation \citep{Fortney2013, OwenWu2013, Jin2014, ChenRogers2016}.  The observed gap therefore also lends credibility to the idea that photoevaporation is a key process that sculpts the population of sub-Jovian class planets, although core powered mass loss---where the luminosity of a cooling rocky core erodes thin H$_2$ envelopes, but preserves thicker ones---is a feasible alternative explanation \citep[e.g.]{Ginzburg2018}.  Subsequent research has also shown that for lower-mass stars, this bimodal distribution for sub-Neptune-sized planets has a gap that shifts to smaller sizes, consistent with smaller stars producing smaller planet cores \citep{Fulton2018}.  These studies also indicate that there are a comparable number of worlds in the terrestrial and mini-Neptune classes, with the proviso that we do not have sensitivity to detect the smallest component of the terrestrial class, or completeness for planets on orbits longer than 200 days.  Nonetheless, Kepler has shown that the Universe is indeed teeming with terrestrial-sized worlds, most of which are likely to have the higher densities associated with terrestrial worlds in our Solar System.     
 
The suite of different types of exoplanets have expanded our knowledge beyond the subset seen in our Solar System.  Fig. \ref{fig:1} shows a schematic for a potential classification scheme for these worlds, based on planetary radius, stellar flux received, and the corresponding boundaries for which condensates will form clouds in these atmospheres \citep{Kopparapu2018}. The planetary radius bins start with a lower limit for terrestrial atmospheric retention \citep{ZC2017}, and end with the radius past which planets transition to brown dwarf stars \citep{ChenKipping2017}, with sub-categories for terrestrials including super-Earths, mini-Neptunes (following the Fulton et al., (2017) distributions), Neptunes, and Jovians.   The incident stellar flux then divides them into hot, warm or cold examples of their size class, with corresponding condensates.  In hot exoplanet atmospheres, ZnS mineral clouds have been considered as possible condensates \citep{Morley2012, Charnay2015}. Moving further away from the star, H$_{2}$O starts condensing in the atmosphere of more temperate worlds.  At lower incident stellar fluxes, CO$_{2}$ and CH$_{4}$ condensates bracket the final boundaries.  In this broader scheme the habitable zone can be thought of as that region between instellations (stellar incident flux) at which liquid water clouds ($\sim 1$ stellar flux) and carbon dioxide clouds ($\sim 0.3$ stellar flux) form (Abe et al. 2011).  Terrestrial planets found within this instellation range are more likely to be habitable than planets in other regions of this diagram.  In the next section, we discuss the calculation of the circumstellar habitable zone in more detail. 

\subsection{Predicting The Habitable Zone}

As discussed in the introduction, the ‘habitable zone’ (HZ) of a star is the circumstellar region where a terrestrial radius and mass planet can maintain liquid water on its surface. In other words, the habitable zone identifies a range of orbital distances where a planet is more likely to support habitable conditions on its surface, and thus can be detectable and characterizable by astronomical observations. 
  While numerous studies have sought to understand the impacts of planetary properties and other factors on the limits of the habitable zone, the most useful region is still likely to be where all habitable zone estimates overlap for a given stellar type, as that will indicate our best understanding of the region of highest probability for surface liquid water.  However, all of these models, despite their commonality in many cases, are still predictions, and based on our understanding of processes working on Earth.  These theoretical predictions of the habitable zone will be subject to revision once observational data on the atmospheric compositions and habitabilty of terrestrial exoplanets are obtained over the next five years to decades. 

Traditionally, one dimensional (1-D) climate models were used to estimate the position of the HZs around different stars \citep{Huang1959,Hart1978,Kasting1993,Selsis2007b,Pierrehumbert2011a,Kopparapu2013a, Zsom2013, Ramirez&Kaltenegger2017}. These models assume Earth-like planets with CO$_{2}$, H$_{2}$O and N$_{2}$ atmospheres, and an active carbonate-silicate cycle, which provides a negative feedback to  buffer atmospheric CO$_2$ as a function of surface temperature \citep{Walker1981}.  This latter assumption will result in ``Earth-like'' planets that have less CO$_2$ than Earth near the inner boundary, and significantly more at the outer boundary of the HZ. The “1-D” nature of the model comes from the atmosphere being approximated as a single column extending from the surface to about $\sim 100$ km in altitude ($\sim 0.1$ millibar), divided into numerous levels where radiative transfer calculations are performed.  Such 1-D column models are meant to capture planet-wide average conditions, in a simple and efficient package. They are often run cloud-free, or with simplistic approximations for the radiative effect of clouds \citep{Kasting1993}, and with atmospheres where H$_2$O and CO$_2$ are the only greenhouse gases.

\begin{figure*}
  \centering
\includegraphics[scale=0.55]{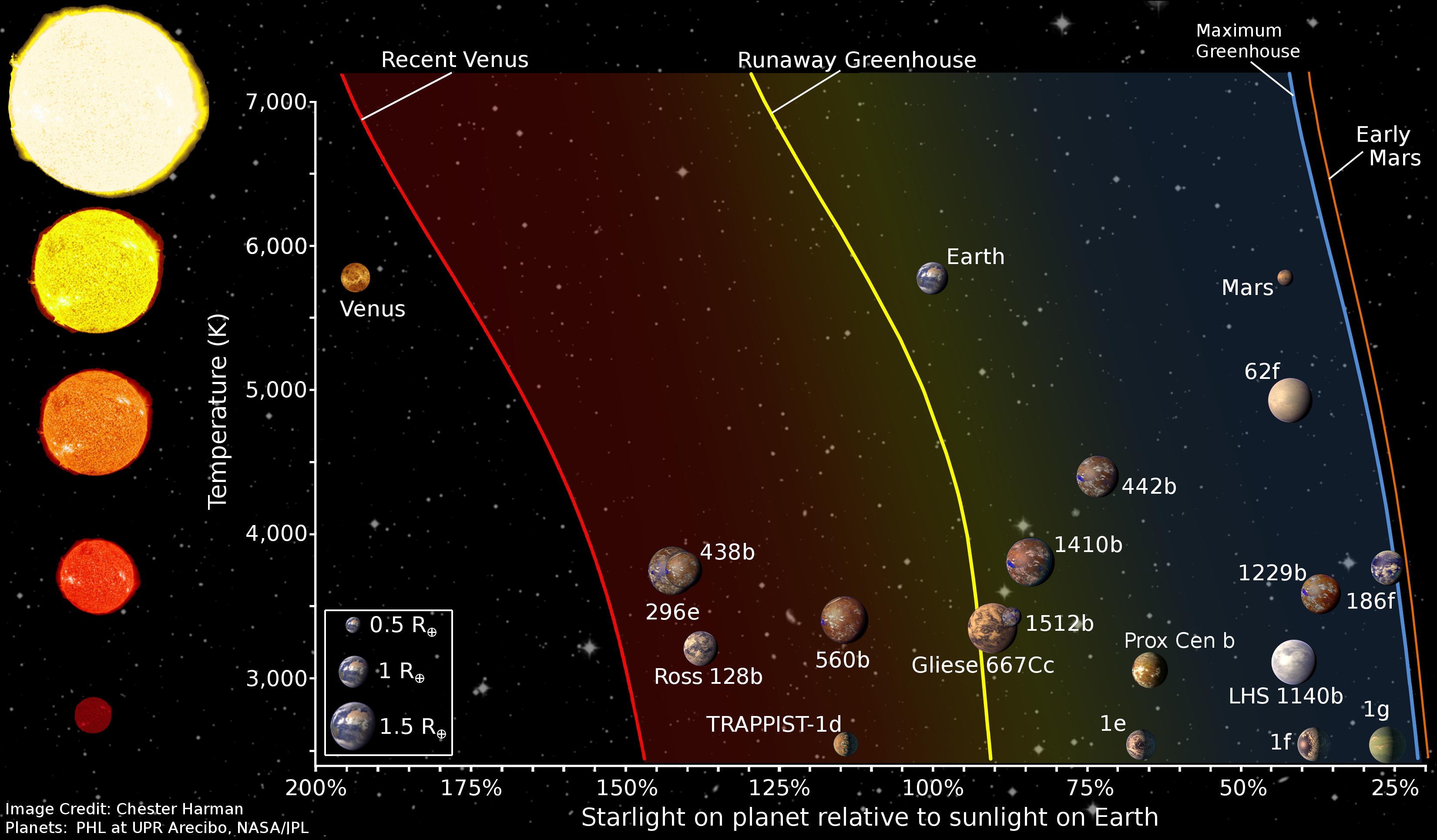}
\caption{\small{\textit{ This figure shows the HZ limits for an Earth-like planet around stars with different stellar temperatures (vertical axis) in terms of incident stellar flux (horizontal axis) on the planet, from a 1-D climate model. The `conservative HZ' is the region between the runaway greenhouse and maximum greenhouse limits. The 'optimistic HZ' is the region between recent Venus and early Mars limits. See text for HZ definitions. Currently confirmed terrestrial exoplanets, along with the Solar system ones, are also shown.}}}
\label{fig:2} 
\end{figure*}

The width of the HZ is defined by inner and outer edges, which are bounded by climate catastrophes. The models simulate where surface liquid water is no longer stable if one pushes the planet closer to the star, increasing the incident stellar flux (inner HZ, IHZ), or away from the star decreasing the stellar flux (outer HZ, OHZ; see Fig \ref{fig:2}. The IHZ proposes two habitability limits: A moist greenhouse limit, where the stellar radiation warms the atmosphere sufficiently so that the stratospheric water vapor volume mixing ratio becomes $> 10^{-3}$ (Earth's H$_{2}$O mixing ratio is $\sim10^{-6}$ at 1 milli bar), causing the planet to lose water by photolysis and then subsequent escape of free hydrogen to space; (2) A runaway greenhouse limit, whereby surface water is vaporized, and the atmosphere becomes opaque to outgoing thermal radiation due to excess amounts of H$_{2}$O in the atmosphere, heating uncontrollably perhaps to beyond ~1500 K \citep{Ingersoll1969, Goldblatt2013}. While the runaway greenhouse is the more violent and catastrophic end, the moist greenhouse is the more proximal.  Habitability could potentially be terminated via the moist greenhouse process long before a thermal runaway occurs. However, some studies found that moist-greenhouse may be inhibited under certain conditions,  because of the sub-saturation resulting in cooler stratospheres \citep{Leconte2013c}. The moist greenhouse limit depends entirely on a planet's inventory of non-condensing gases. A planet with negligible N$_{2}$ and Ar would enter the moist greenhouse limit even if it were in a Snowball state \citep{Wordsworth2014}. 

The OHZ is defined by the “maximum” greenhouse limit, where the warming provided by the build up of atmospheric CO$_{2}$ (due to the active carbonate-silicate cycle) is maximum. Models indicate that this occurs with $\sim6$ to 10 bars of CO$_{2}$ in the atmosphere. For thick CO$_{2}$ atmospheres, the enhancement of the greenhouse effect from adding more CO$_{2}$ begins to saturate, while the reflectivity of the atmosphere due to Rayleigh scattering from a thick atmosphere continues to increase.  Beyond a certain amount of atmospheric CO$_{2}$, increases in scattering win out over the increases to the greenhouse effect, causing the planet to experience cooling instead of warming.  This turning point marks the maximum CO$_{2}$  greenhouse outer edge limit to the HZ. 

To guide our search for liquid water on a planetary surface, the conservative habitable zone uses the runaway greenhouse inner limit and the maximum greenhouse outer limit, but a more optimistic habitable zone can be defined empirically based on phenomena in our Solar System.  For the optimistic IHZ one can define a “recent Venus” limit, based on geological evidence that Venus has not had liquid water on its surface for at least the past 1 billion years (Solomon \& Head 1991).  If we assume Venus was habitable right up until 1 billion years ago, then the recent Venus limit is the equivalent distance from our modern Sun that would have matched the insolation at Venus 1 billion years ago under a fainter Sun. For the outer edge, there is a corresponding “early Mars” empirical estimate, based on geological evidence that suggests that Mars had liquid water on its surface 3.8 billion years ago. These optimistic empirical limits, and the conservative limits calculated using climate models can be used as a first order means of identifying habitable planet candidates for follow-up observations. All the currently known terrestrial exoplanets that are in their host star’s HZs are shown in Fig. 2. 

Although 1-D climate models are relatively fast to run and  can include reasonable physics and chemistry (\cite{Lincowski2018}, 
there are instances where complex 3-D climate models are particularly needed to understand the impacts of planetary circulation, rotation rate and cloud formation on climate and habitability.  In particular 3D GCMs can address impacts of ice-albedo feedback  \citep{Joshi1997, Joshi2003, Shields2013} planetary volatile abundance, circulation and climate \citep{Abe2011, Wordsworth2010, Edson2011, Pierrehumbert2011a, Pierrehumbert2016, Way2016} the impacts of cloud formation on the inner edges of the habitable zone \citep{Leconte2013c, Yang2013, Yang2014b, Wolf2015a, Way2015, Godolt2015, Kopparapu2016, Kopparapu2017, Haqq2018}   and the effect of feedbacks betweeen ice formation, atmospheric composition and surface temperature for the outer edge \citep{Turbet2017a, Turbet2017b}. 

Estimates of the HZ for Sun-like stars (F,G,K dwarfs) from 3-D climate models are within $~\sim5$ to $7\%$ of the predictions of 1-D models. However, 3-D models for ocean-covered planets orbiting late K and M dwarf stars predict significantly expanded HZs, which is due in part to planetary dynamical spin states.  Planets near and within the IHZ of these later-type stars are close enough to the star that tidal locking is a likely dynamical outcome \citep{Ribas2016}. If the planet's orbital eccentricity is small, this can result in synchronous rotation, where the rotation period of the planet equals its orbital period \citep{Dole1964, Peale1977, Dobrovolskis2009, Leconte2015, Barnes2017}, producing permanent day and night sides.  Tidally-locked planets are more likely to have slower rotational periods than the Earth. This slower rotation diminishes the atmospheric Coriolis force and changes atmospheric circulation,  affecting relative humidity, clouds, heat transport and ultimately the climate.  On more rapidly rotating planets, like the Earth, the Coriolis force deflects air parcels to the right in the northern hemisphere, and to the left in the southern hemisphere, producing latitudinally-banded cloud patterns with a cloudy (reflective) equator and clearer sub-tropics. However, for slowly rotating planets the Coriolis force is too weak, and instead strong and persistent convection occurs at the sub-stellar region, creating a stationary and optically thick cloud deck. This causes a strong increase in the planetary albedo, cooling the planet, and stabilizing the climate against a thermal runaway for large incident stellar fluxes. Thus, an ocean covered planet may be able to maintain clement global-mean surface temperatures ($\sim$280 K) around M-dwarf stars at much higher stellar fluxes than predicted by 1-D models. This in turn extends the inner-edge of the HZ closer to the star, increasing the width of the HZ \citep{Yang2013, Yang2014b, Kopparapu2017}.

Modeling to better understand limits at the outer edge of the habitable zone have also been undertaken.  GCM simulations indicate that planets at the outer edge of the HZ around M-dwarfs are less susceptible to snowball climates due to the lower snow/ice albedo at near-IR wavelengths, which, interacting with the M dwarf's red/NIR incident spectrum, which causes surface ice to melt more easily compared to under a Sun-like incident spectra \citep{Shields2013}.  This may extend the outer edge of the HZ around M-dwarfs to lower stellar fluxes compared to models that do not include the ice-albedo feedback. However, oscillations between ice-free and globally glaciated states, called “limit cycles”, could occur on planets with volcanic outgassing rates that are too low to sustain a CO$_{2}$-warmed climate \citep{Kadoya2014, Kadoya2015, Menou2015, Haqq2016}.  Planets orbiting Sun-like stars, and F-stars in particular, may be more susceptible to limit cycles due to the stronger ice-albedo feedback with the strongly blue spectrum of the F-star, reducing the extent of the HZ for Sun-like stars.  However, for cases where volcanic outgassing is more pronounced, and/or the atmosphere includes various cocktails of other greenhouse gases such as H$_{2}$ and CH$_{4}$ in addition to high amounts of CO$_{2}$, then the OHZ may be extended \citep{Pierrehumbert2011b, Wordsworth2013c, Seager2014, RKalt2017}. 
Determining which of the above processes, if any, actually govern climates near the edges of the habitable zone awaits near-term observations of terrestrial atmospheres with JWST and ground-based telescopes that may be able to identify greenhouse gas compositions and search for signs of runaway greenhouse processes or ocean loss \citep{Morley2017, Lincowski2018, Lustig-Yaeger2019, Turbet2019}

\subsubsection{Habitable Zones Around Binary Stars}

Single stars are the primary focus in the search for habitable planets, but exploration of the habitable zone for binaries is being undertaken, in anticipation of the eventual discovery of terrestrial planet candidates orbiting binary stars. Binary stars are common, with nearly half of all Sun-like stars residing in binary (and higher multiple star) systems.  
At the time of writing there are six confirmed planets orbiting one member of a sub-20 AU binary stellar system (i.e. circumprimary planets or S-type systems, \citet{Kley2014}) and 12 confirmed planets orbiting within 3 AU of both members of sub-AU binary star systems (circumbinary planets or P-type systems, e.g. \citet{Welsh2015,Kostov2016}). Based on known circumbinary systems, estimates suggest a $1-10\%$ occurrence rate of Neptune- to Jupiter-sized planets (e.g. \citet{Armstrong2014,Kostov2016} WanWelsh et al. 2015). Almost half of known circumbinary planets (planets orbiting both the stars of a binary stellar system) reside in the HZ \citep{Doyle2011, Orosz2012b, Orosz2012a, Welsh2015, Kostov2013, Kostov2016}, but these planets are not terrestrials and so are likely not habitable.  Discovering transiting planets orbiting binaries is challenging, and is usually done by eye, because their transits are strongly aperiodic.  Promising new techniques are being developed to automate the search, and increase our chances of eventually finding smaller terrestrial planets \citep{Windemuth2019}.

Meanwhile, there has been some progress in predicting the HZs of Earth-like planets around binary stars \citep{KH2013, HK2013, Eggl2012, Eggl2013, Forgan2014, KaneHinkel2013, Forgan2016, PoppEggl2017, WangCuntz2019}, although this is also challenging.  Stellar insolation recieved by habitable zone planets in circumbinary systems can change by up $50\%$ due to the oscillations of the host stars.  Theses changes in stellar insolation occur on timescales of $\sim 10$s to $\sim100$ Earth days, and can drive extreme weather and seasonality on circumbinary planets.  This could affect prospects for habitability on such worlds. Currently, there are efforts by various groups to simulate such systems using hierarchical models of 1-D, energy balance model (EBM) and GCMs.

\subsection{Occurrence Rates for Potentially Habitable Worlds}

    Now that we understand the size range most likely associated with a terrestrial-type world, and have a working estimate for the limits of the habitable zone, we can identify those known planets that are more likely to be habitable (see Fig. \ref{fig:2}), and calculate initial estimates of the occurrence rate of potentially habitable planets (likely terrestrials in the habitable zone).  This quantity, $\eta_{\oplus}$, is defined as the fraction of stars that have at least one planet in the HZ. Current estimates of $\eta_{\oplus}$ from the {\it Kepler} data for Sun-like stars range from $0.22 \pm 0.08$ \citep{Petigura2013} to $0.36 \pm 0.14$ \citep{Mulders2018}, with some estimates as low as $0.02$ \citep{Foreman-Mackey2014}. However, realizing this large variation, and to help achieve a useful community-wide consensus of occurrence rates for FGK stars, the NASA funded Exoplanet Exploration Program Analysis Group (ExoPAG) led Study Analysis Group 13 (SAG13). SAG13 standardized a grid of period and planet radius, interpolated most published occurrence rates to this grid, collected additional contributions from the community, and compiled the results to form a community-wide average with uncertainties. Integrating the SAG13 occurrence rates over the boundaries of stellar and planetary parameters that the community agreed to, gives $\eta_{\oplus} = 0.24^{+0.46}_{-0.16}$ \citep{Stark2019}. This value of $\eta_{\oplus}$ was used to calculate the FGK exo-Earth yields for mission concept studies LUVOIR and HabEX. For M-dwarf stars
$\eta_{\oplus}$ is estimated to be
$0.16^{+0.17}_{-0.07}$ for conservative HZ, and $0.24^{+0.18}_{-0.08}$ for the optimistic HZ \citep{Dressing2015}.


Another potential source of habitable worlds are exomoons of HZ Jovian planets.  Since the launch of Kepler telescope, exomoon candidates have received increased attention \citep{Kipping2012, Szabo2013, Simon2015, Agol2015}. Particularly, the habitability of exomoon candidates has been explored both with theoretical models \citep{Heller2012, HinkelKane2013, ForganKipping2013, HellerBarnes2013, HellerBarnes2015, Lammer2014, ForganDobos2016, Dobos2017, HaqqHeller2018} and observational efforts to discover exomoon candidates in the HZ of their host stars \citep{Kipping2013, Kipping2014, Kipping2015, Forgan2017}. Recent occurrence rate estimates of giant planets (3 - 25 R$_{E}$) within the optimistic HZ of Kepler stars find a frequency of $(6.5 \pm 1.9)\%$ for G stars, $(11.5 \pm 3.1)\%$ for K stars, and $(6 \pm 6)\%$ for M stars \citep{Hill2018}. If one assumes that each giant planet has one large terrestrial moon, then these moons are less likely to exist in the HZ than terrestrial planets. However, if each giant planet holds more than one moon, then the occurrence rates of moons in the HZ would be comparable to that of terrestrial planets, and could potentially exceed them. Although there is no robust detection of exomoons presently, there are tentative detections \citep{TeacheyKipping2018, Rodenbeck2018} indicating that a confirmed exomoon discovery is imminent. 

\subsection{Moving Beyond the Habitable Zone: Factors Affecting Habitability} 

Although the HZ provides an excellent first-order means to quickly assess the potential habitability of a newly discovered planet, and will be even more powerful if observationally confirmed, there is a growing realization that the habitable zone is in many ways too simplistic.  Although it provides a zeroth order assessment of whether the planet may be able to support liquid water on its surface now, it does not take into account the formation and subsequent evolution of the planet, or the diversity of characteristics or interactions between planet, star and planetary system that can shape whether or not the planet was able to acquire or maintain liquid on its surface.  Planetary habitability is now recognized as the interdisciplinary,  multi-factorial outcome of a planet's evolution and planetary system environment. 

\begin{figure*}
\centering
\includegraphics[scale=0.9]{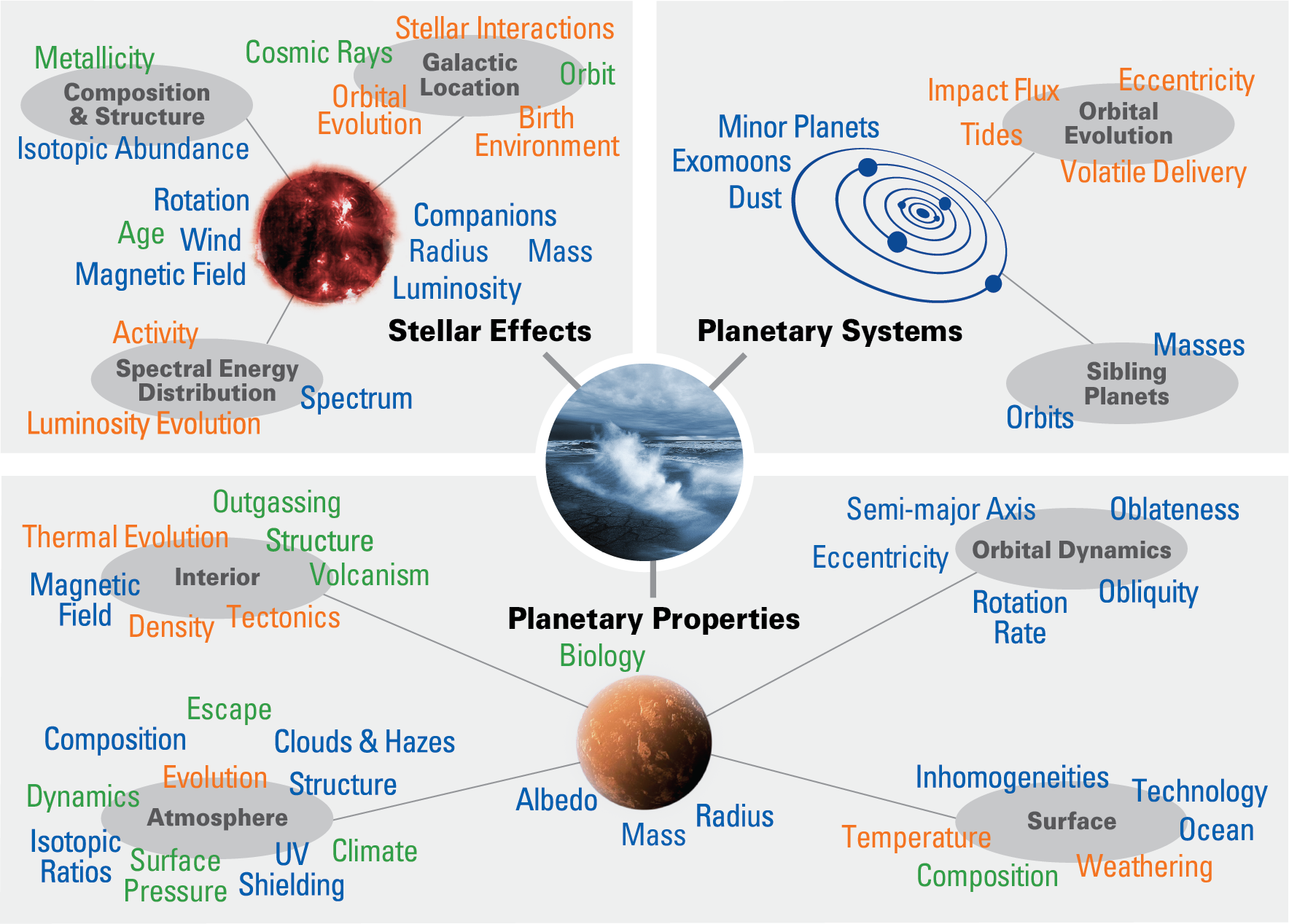}
\caption{\small{\textit{Factors Affecting Habitability. This diagram shows currently understood planetary, stellar and planetary system properties that may impact planetary habitability. The larger the number of these factors that can be determined for a given habitable zone candidate, the more robust our assessment of habitability will be. Font color denotes characteristics that could be observed directly with sufficiently powerful telescopes (blue), those that require modeling interpretation, possibly constrained by observations (green), and the properties or processes that are accessible primarily
through theoretical modeling (orange). From \cite{MeadowsBarnes2018}}}}
\label{fig:3} 
\end{figure*}

Factors---characteristics and processes---that impact habitability can be identified in three major areas: Planetary characteristics, stellar characteristics, and planetary system characteristics.  Habitability is influenced by these properties, but also by the interactions that occur between these components as a function of time, that allow a planet to acquire and maintain liquid water on its surface (Fig. \ref{fig:3}).  While each of these factors is important, only a subsample are potentially observable (denoted by the blue text in Figure \ref{fig:3}), and so could be used in the near-term to help characterize habitability. Nonetheless, continued theoretical work into understanding how each of these factors impacts habitability will better prepare us to search for habitable planets and life, and to interpret upcoming data on terrestrial exoplanet environments.  


\section{PLANETARY CHARACTERISTICS FOR HABITABILITY}

The planet's environment, it's mass, radius, orbit, interior, surface and atmosphere set the stage for habitability.  Once life has evolved on a habitable world, it becomes a planetary process that can also impact its environment \citep{Lovelock1974, Goldblatt2009,Tziperman2011, Lenton2012}.  A detailed discussion of life as a planetary process on Earth, and its coevolution with our environment over Earth's history can be found in Chapter 4 of this book. Below we describe several of the key planetary characteristics and processes that support habitability, and briefly describe our likely ability to observe these characteristics on exoplanets.

\subsection{Effect of Mass and Radius on Habitability}

While true limits on planetary radii or masses for habitable planets are currently unknown, the radius/mass range within which a planet is more likely to be habitable can be constrained.  As discussed above, observations of small \textit{Kepler} planets, for which the mass, radius and density are known, have suggested that 1.5 R$_\oplus$ radii is the upper limit for an exoplanet to be more likely to have a predominantly rocky composition \citep{Weiss2016,Rogers2015,Fulton2017}. Above this limit, planetary densities drop significantly, suggesting rocky cores with thick hydrogen envelopes, mini-Neptunes, which would be much less likely to be habitable \citep{Owen2016}. The lower mass limit for which a planet is likely to have sufficient radiogenic heating to drive plate tectonics and atmospheric replenishment via outgassing has been theoretically calculated as 0.3 Earth masses for an Earth-like composition (0.7 Earth radii for an object of Earth's density) \citep{Raymond2007,Williams1997}.

A planet's mass impacts planetary habitability in multiple ways.  It provides radiogenic heating from long-lived radionuclides to drive internal heating and tectonics \citep{Lenardic2012}, as well as generation of a magnetic field \citep{Driscoll2015}, which is a key parameter that determines atmospheric retention  \citep{Chassefiere2007,Lammer2012, Egan2019}.  Planetary mass, via planetary gravity, also controls atmospheric scale height, which can change the rate the planet radiates to space, and modify its climate and the limits of the HZ  \citep{Kopparapu2014}.  Mass is also a key parameter in atmospheric retention, which is dependent on the interplay of planetary mass, radius and insolation \citep{Zahnle2017}. Even though Mars lies within the HZ, at 0.1 Earth masses it has not been able to retain or replenish a sufficiently large atmosphere to maintain liquid water on its surface, and so is below the habitable mass limit.  If the locations of Venus and Mars were swapped, it is possible that the Solar System might have supported two habitable planets. 

Planetary radii and masses can be relatively straightforward to measure, depending on the technique used to detect the planet. Exoplanet radius is straightforward to measure if the planet transits and the stellar radius is well-known, since planet size is derived from the drop in measured flux as the planet passes in front, and the size of the star (Borucki et al. 2010; Batalha et al. 2011).  Size is extremely
challenging to observe if the planet does not transit, due to an inherent size-albedo degeneracy at visible-NIR wavelengths, that can be broken with observations in the thermal infrared (Des Marais et al. 2002). However,  exoplanet masses can be measured using transit timing variations (Deck and Agol 2015; Agol and Fabrycky 2017), astrometry (Benedict et al. 2006), or radial velocity measurements (Mayor
and Queloz 2012) combined with planetary system inclinations derived from transit duration observations (Borucki et al. 2010) or high-resolution spectroscopy measurements of exoplanet orbital velocity (Snellen et al. 2010; Luger et al. 2017a).

\subsection{Planetary Orbit, Obliquity, and Rotation Rate}

The planetary orbital parameters, such as semi-major axis, eccentricity, obliquity, and rotation rate, have a significant influence on planetary habitability through their control on the stellar radiation received by a planet over its orbit, and associated feedbacks on the climate system.  Fundamentally, the time-averaged amount of stellar radiation received by a planet is determined by its distance from the host star, and thus its semi-major axis.  Stellar radiation is the primary source of energy for planetary atmospheres.  First-order assessments of planetary habitability and the habitable zone typically rely on the received stellar flux as a primary metric (Fig. 1 and Fig. 2).  However, the combination of eccentricity, obliquity, and planet rotation rate contribute to complicated temporal and spatially dependent variations of the stellar radiation received by a planet \citep{Berger1993, Shields2016b}.  Orbital system parameters may also evolve over time \citep{Armstrong2014}. These characteristics of orbital systems can affect the prospects for habitability, sometimes strongly.

Planets on eccentric orbits receive significant variations in stellar radiation over the course of their orbits as the star-planet distance changes between aphelion and perihelion.  If the eccentricity is large, this can result in seasonal changes to a planet's surface temperature.  However Earth-like planet's, those with significant oceans and atmospheres, have a large thermal inertia and thus can buffer time-varying changes in the stellar radiation.  The long-term climate stability of eccentric planets is determined primarily by the average stellar flux received over the course of their orbit, and not by the extremes received at aphelion and perihelion respectively \citep{Williams2002,Bolmont2016, Dressing2010,WayGeorgakarakos2017, Adams2019}.  Still, seasonal temperature extremes and fluctuating environmental conditions could pose significant challenges for the evolution and adaptation of complex life \citep{SherwoodHuber2010}. Generally, planets with thicker and wetter atmospheres are better able to buffer time-dependent changes in stellar radiation, while planets with thinner and drier atmospheres will be more susceptible climate oscillations driven by time-varying stellar radiation.  

Earth has a small but non-zero eccentricity.  However Earth's seasons are not driven by its eccentricity, but rather by its obliquity.  Obliquity is the axial tilt of the planet's rotational axis relative to the star-planet plane.  A planet's obliquity determines the latitudinal variation in the received stellar radiation.  As the planet orbits the star, a non-zero obliquity results in a meridional migration of the sub-stellar point north and south of the equator.  Still, for low obliquity planets, like Earth, the annually averaged stellar radiation remains centered at the equator, with meridional excursions of the sub-stellar point creating the seasons.  However, for high obliquity planets ($>$54 degrees), the time-averaged pattern of the stellar radiation reverses, with a maximum flux received at polar regions and a minimum at the equator \citep{Jenkins2000}.  This peculiar pattern of stellar radiation creates unique climate states where ice belts may accumulate around the equator while the poles remain temperate and habitable \citep{Spiegel2009,Spiegel2010,Kilic2017}.  For high obliquity planets bistability thresholds between habitable temperate climates and uninhabitable snowball climates are notably altered compared to thresholds identified for low obliquity worlds \citep{Linsenmeier2015, Rose2017, Colose2019}.  

Planetary rotation rate controls the diurnal period (the length of day).  Uniquely among orbital parameters, the planetary rotation rate imparts a significant impact on the circulation state of the atmosphere through the action of the Coriolis effect.  The Coriolis effect is a fictitious force which arises due to Earth's rotation, and deflects large scale atmospheric motions relative to Earth's surface.  Changes to planet's rotation rate, and thus Coriolis effect, can trigger a different atmospheric circulation regimes to emerge \citep{Carone2015, Noda2017,Haqq2018}.  The atmospheric circulation state affects horizontal heat transport and the spatial distribution of clouds, each of which can significantly affect the climate and habitability of a planet \citep{Yang2013,Kopparapu2017,Wolf2019, KomacekAbbot2019}. 

For slowly rotating planets, like Venus or planets found around M dwarf stars where tidally-locking is expected, the Coriolis effect is weak and the atmospheric circulation regime  supports the creation of thick stationary clouds at the sub-stellar point which effectively reflect sunlight and permit a planet to be habitable at higher insolation levels \citep{Yang2013, Kopparapu2016, Way2016}. For more rapidly rotating planets, like the Earth, a stronger Coriolis effect leads to predominately zonal circulation and the creation of zonally banded cloud decks which are less efficient at reflecting sunlight \citep{Yang2014a, Kopparapu2017}.

For planets that are tidally-locked and synchronously rotating (in a 1:1 spin-orbit resonance), one side of the planet always faces the star, and the other side of the planet is in permanent darkness.  On such worlds, the day-night temperature differences can be enhanced leading to the possibility of the atmosphere freezing out or ``collapsing'' on to the night side \citep{Joshi1997,Joshi2003,Turbet2016, Leconte2013b}.  This is true especially for colder and thinner Mars-like atmospheres, however modeling has shown that thicker Earth-like atmospheres can maintain sufficient day-to-night heat transport to prevent collapse \citep{Joshi1997, Wordsworth2015, Kopparapu2016, Wolf2019}.  Note that synchronous rotation is not guaranteed for tidally-locked planets around M dwarf stars, however, as trapping into spin-orbit resonances (like Mercury's) are also possible \citep{Hut1981,Rodriguez2012,Ribas2016}. A large atmosphere may also prevent synchronization \citep{Gold1969,Leconte2015}. The tidal damping of the rotation rate into a synchronous state is model dependent \citep{FerrazMello2008,Barnes2017} and depends on the planet's structure \citep{Henning2014} and, if present, the tidal dissipation in a planet's ocean \citep{Egbert2000,Green2017}.  

Planetary orbital properties are generally amenable to observations.  Semi-major axes can be observationally determined using transit, radial velocity and astrometry.  Eccentricity is also straightforward to measure with radial velocity and astrometry but more challenging with transit, and in the latter case more accurate if both the primary and secondary eclipse can be observed \cite{Demory2007}.  However, determining obliquity and rotation rate for a terrestrial planet will require time-dependent mapping using direct imaging observations \citep{Fujii2017,Lustig2017,Kawahara2010,Cowan2009}.  By observing the thermal emission as a function of orbital phase, different climate states may be discernable \citep{Leconte2013b, Yang2013, Haqq2018, Wolf2019, Adams2019}

\subsection{Planetary Interior and Geological Activity}

The interior of a planet plays a critical role in determining the habitability of a planet.  An active and dynamic interior can drive the generation of a magnetic field \citep{Olson2006,Driscoll2013} and outgassing \citep{Driscoll2014}, which are key for producing and maintaining a secondary atmosphere. 

The atmospheres of terrestrial planets in our solar system are a product of outgassing. Primordial atmospheres, if they ever existed for terrestrial planets, are dominated by H$_{2}$ and He with traces of Ar and Ne accumulated during the formation of the Solar system  from the gaseous nebular disk \citep{Sekiya1980, Lammer2018}. However, such an atmosphere will have only a fleeting existence on a newly formed terrestrial planet, because the temperatures are hot enough for the lighter gases (H$_{2}$ and He) to escape the low gravity well of the planet (a simple relation between mean kinetic energy and the internal energy due to the temperature of the gas).  The more heavy icy materials (H$_{2}$O, CH$_{4}$, NH$_{3}$), on the other hand, combine with the rocky materials (like iron, olivine) and get integrated into the crust and the mantle. If the terrestrial planet is big enough to maintain the formation heat, it can sustain an active tectonic activity, which results in volcanism, which in turn releases these trapped icy materials producing secondary atmospheres.

Tectonic activity on a planet can influence the habitability of a terrestrial planet through cycles of volcanic outgassing and consequent weathering of the released gases. Tectonic activity also creates weatherable topography which has a long-term impact on the evolution of the climate \citep{Lenardic2016b} in terms of negative feedback between silicate weathering (the loss process for atmospheric CO$_{2}$) and surface temperature \citep{Walker1981}. Furthermore, tectonic activity similar to that of the Earth facilitates an efficient water cycling between the surface and the interior, sustaining oceans on the planet \citep{Sandu2011, Cowan2014, Schaefer2015, KomacekAbbot2016}. Tectonic and volcanic activity cycles that operate beyond the characteristics of the Earth may occur on exoplanets of varying mass and composition, such as stagnant lid \citep{Solomatov2000}, episodic tectonics \citep{Lenardic2016b} and heat pipe \citep{Moore2013,Moore2017b}. On the other hand, coupling melt models \citep{Katz2003} with mantle convection models can simulate volcanic outgassing, which depends upon the composition and the internal temperature of the planet.

Augmenting a planet's primordial internal heat for extended periods of time can be possible by radionuclide decay \citep{Dye2012} or tidal stress \citep{Jackson2008,Barnes2009a}.
Radionuclide decay releases high energy particles, which can be absorbed in the planetary interior. The nature of a planet's radiogenic sources is predetermined during the formation of the system. Different radio isotopes decay at different rates. For example, $^{26}$Al is a short-lived isotope whose half-life is just $\sim$700,000 years, which is suspected to have been present during the Solar System’s formation. $^{26}$Al is produced in supernova explosions of massive stars, which provide the ingredients and the initial `fuse' (through shock waves) for forming a planetary system. Short-lived isotopes drive the differentiation of elements inside a planet through their radiogenic heating during the primordial stages of their formation. Once the differentiation begins, the frictional heat of partitioning elements sustains the internal heat deep inside the planet.

On the other hand, $^{40}$K is a long-lived radio isotope with a half life of more than a billion years old. Long-lived isotopes generally concentrate near the crust and the mantle, providing heat at these layers, due to their large size preventing dense packing deep within the Earth. 

The internal heat energy needed for tectonic activity can also be generated  by tidal heating, where the differential gravity of the planet due to a companion (usually the host star or nearby planet) causes internal stress, and energy is deposited by friction \citep{Jackson2008,Driscoll2015}.  
A key area of future research for the impact of planetary interiors on terrestrial planet evolution and habitability will be understanding degassing from terrestrial planets of different composition, including the potentially volatile-rich migrated terrestrial planets found orbiting M dwarfs \citep{Gillon2017,Luger2017c,Grimm2018}.

Planetary interior properties will be challenging to determine observationally.  However, precise characterization of the planetary system's orbital state could theoretically be used to yield constraints on planetary interior structure---including determination of the rigidity of the planetary body and its susceptibility to tidal deformation \citep{Buhler2016,Becker2013}.  Constraints on the planet's interior structure and composition could also be gleaned from a combination of knowledge of the star's composition, the planet's mass and radius, and planet formation models \citep{Dorn2015,Unterborn2016}.  For multi-planet transiting systems, density measurements to constrain interior composition could also be obtained from observations of planetary radius from transit, and planetary mass from radial velocity, astrometry or transit timing variations. Hints of the  planet's interior composition may also be obtained from measurements of atmospheric and cloud composition, which may point to a  steady-state volcanic outgassing source, as it does for the clouds of Venus \cite{Bullock1997}. Transmission or direct imaging observations may also reveal transient compositional changes in atmospheric gases \citep{Kaltenegger2009,Kaltenegger2010} or aerosols \citep{Misra2015} that are indicative of ongoing volcanic activity. 

The redox state of a planet also influences its habitability. The escape of hydrogen is important as it can drive water loss, and can also alter the planet’s surface redox state to more oxidizing \citep{Catling2001, Catling2005, Kump2008,  Armstrong2019}.   Similarly, the degree of iron segregation to the core and/or iron redox disproportionation sets the redox state of the mantle \citep{Frost2008}, which determines whether reducing or oxidizing gases get outgassed by volcanoes.  These gases have a large impact on the composition of the secondary terrestrial planetary atmosphere \citep{Wordsworth2013b}.  Highly reducing conditions have climate consequences but they are also fundamental to driving prebiotic chemistry, particularly when abundant HCN is present \citep{Ferris1984, Orgel2004}.
Conversely, highly oxidizing surface conditions are not only a roadblock to the origin of life (see the chapters by Hoehler et al. and Baross et al. in this volume), they can also make conditions toxic for complex life altogether if O$_{2}$ levels are high enough \citep{Baker2017}.

\subsection{Magnetic Fields}

 Magnetic fields are an important factor when considering the   habitability of a planet, as they  may protect planets from losing volatiles (such as water) through stellar wind interactions  \citep{Chassefiere2007,Lundin2007,Lammer2012, Driscoll2013, donascimento2016, Driscoll2018}. However, this ``magnetic umbrella'' hypothesis is still debated, as the magnetic field may also increase the interaction area with the solar wind, which could drive increased escape \citep{Brain2013, Egan2019}.  Although {atmospheric escape} was often assumed to be due to thermal processes, and independent of the magnetic field  \citep[][e.g.]{Hunten1976,Watson1981,Lammer2008} modeling had suggested that the magnetic-limited escape rate does indeed decrease with increasing planetary magnetic moment \citep{Driscoll2013}.  However, recent modeling suggests that the situation is more complex, and that whether a magnetic field decreases or increases atmospheric escape is dependent on multiple factors including the strength of the planet's intrinsic magnetic field and the incoming solar wind pressure \citep{Egan2019}.        
  
  Convection in the iron-rich core maintains the planetary magentic field, and is tied to the interior thermal evolution which can reveal the energetic state and history of the interior \citep{Stanley2010,Stevenson2010,Schubert2011}.  A magnetic dynamo is generated via convection in the outer core \citep{Olson2006}. This process is influenced by the planetary rotation rate and core material properties, and enhanced by buoyancy driven by the core cooling rate, which is in turn controlled by the overlying mantle. Plate tectonics cool the planet's interior, and Venus' lack of plate tectonics may explain its lack of a magnetic field \citep{Nimmo2002}, although as a counterpoint Mercury and Ganymede maintain dynamos below their stagnant lids \citep{Ness1978,Kivelson1997}.  Observations and models of {volatile loss} rates for planets and satellites with or without magnetic fields will provide additional insight into the generation of magnetic dynamos and the extent of magnetoprotection of atmospheres. Detecting the presence of a magnetic field on an exoplanet will be challenging, however recent observations have inferred magnetic fields around hot-Jupiters \citep{Cauley2019}. Magnetic star-planet interactions involve the release of energy stored in the stellar and planetary magnetic fields. These signals thus offer indirect detections of exoplanetary magnetic fields. Large planetary magnetic field strengths may produce observable electron cyclotron maser radio emission by preventing the maser from being quenched by the planet's ionosphere \citep{Ergun2000}. Intensive radio monitoring of exoplanets will help to confirm these fields and inform the generation mechanism of magnetic fields. 
 Constraints on magnetospheric strength might be gained in the near future with the detection of auroral lines in high-resolution spectra of exoplanet such as Proxima Centauri b \citep{Luger2017a}, although caution will be needed in discriminating these from more diffuse, globally-prevalent airglow lines.  
Radio emission may also indicate a planetary magnetic field,  with coherent emission frequency providing a constraint on the strength of the field itself \citep{Zarka2007,Hess2011} and characteristic radio emission from the star due to its interaction with a magnetized planet \citep{Driscoll2011,Turnpenney2018}.
 
\subsection{Atmospheric Properties}

While the atmosphere of a terrestrial planet generally constitutes only a minuscule fraction of the planet's overall mass and radius, atmospheric properties play an out-sized role in determining habitability.  For example, while the Earth and Moon each receive the same amount of stellar radiation, their surface environments are decidedly different because Earth has an atmosphere and the Moon does not.  An atmosphere is an envelope of gas that surrounds a planet, and is retained due to the force of gravity.  An atmosphere is a necessary condition for a planet to have liquid water at its surface because water can only remain stable as a liquid under a relatively narrow range of temperatures at a given pressure.  Without adequate pressure, surface liquid water would irreversibly evaporate or sublimate away, as it would on Mars and the Moon despite each residing within the habitable zone \citep{Kopparapu2013a}.  Adequate atmospheric pressure is particularly important for synchronously rotating planets where day-night temperature differences can grow large, potentially resulting in atmospheric collapse onto the night-side \citep{Wordsworth2015, Turbet2016, Turbet2017}.  However atmospheric collapse can be countered by denser atmospheres which promote efficient heat transport.  Of course, atmospheres that are too dense may negatively impact habitability, by becoming opaque to stellar radiation.  This may be particularly problematic for planets that do not lose their primordial H$_2$ atmospheres \citep{Owen2016}, and for habitable planets near the outer edge of the habitable zone which require thick atmospheres to stay warm.

Appropriate surface temperatures for liquid water are maintained through a delicate balance between absorbed incoming radiation from the star, emitted thermal radiation from the planet, and horizontal heat transports \citep{Trenberth2009}.  The constituents of planetary atmospheres, including gases, clouds, and aerosols, critically modulate a planet's energy balance and thus its climate and ultimately its local surface temperatures \citep{Read2015}.  Maintaining the right surface temperatures for liquid water to exist requires having just the right amount and combination of atmospheric gases. CO$_2$ is perhaps the most familiar greenhouse gas, and helps maintain clement temperatures on Earth.  The longterm regulation of CO$_2$ in the atmosphere via the silicate weathering cycle is thought to keep planets habitable despite large differences in their received stellar flux \citep{Walker1981}.  Planets at large distances from their star may be kept habitable by the strong greenhouse effect provided by several bars of CO$_2$ \citep{Kasting1993, Kopparapu2013a, Selsis2007a}.  However, as the atmosphere becomes thick, Rayleigh (molecular) scattering increases, reflecting stellar energy away from the planet.  Other greenhouse gases may also help keep planets sufficiently warm in the habitable zone, including CH$_4$, N$_2$O, NH$_3$, and also collision-induced absorption of N$_2$ and H$_2$ \citep{Wordsworth2013b, Pierrehumbert2011b, Ramirez&Kaltenegger2018, Koll2019}.  

With respect to the climate of habitable planets, H$_2$O is perhaps the most interesting atmospheric constituent, and not because it is a prerequisite for life.  On a robustly habitable planet, like Earth, the expected surface and atmospheric temperature variations allow water to exist in all three thermodynamic phases simultaneously in the atmosphere, oceans, and on the surface.  Each phase of water contributes strong competing feedbacks on a planet's climate.  Water vapor is a strong greenhouse gas and near-infrared absorber and acts to warm a planet.  High-altitude ice water clouds (i.e. cirrus clouds) act also as strong greenhouse agents and warm a planet.  Liquid water clouds (e.g. stratus clouds) are highly reflective, raising the albedo and cooling a planet.  Finally, water that condenses on the surface as snow and ice also raises the albedo and cools a planet.  The water vapor greenhouse feedback and ice-albedo feedbacks are both positive climate feedbacks, meaning that they will amplify climate perturbations, potentially leading to climate catastrophes of a runaway greenhouse and runaway glaciation and the end of habitability.  While water is of course critical for the existence of life, water has an inherently destabilizing force on the climate system.

Beyond the regulation of climate, atmospheres play other important roles that factor into habitability.  For instance, for oxygen-rich planets, stratospheric ozone plays a major role in maintaining surface habitability by shielding harmful UV fluxes from reaching the surface \citep{Segura2003, Segura2005, Rugheimer2015}.  Alternatively, anoxic planets may form Titan-like photochemical hazes that are created in the upper atmosphere and also constitute a significant UV shield that could protect life on the surface \citep{Sagan1997, Wolf2010, Arney2016, Arney2017}.  Although, while O$_3$ has little effect on a planet's surface temperature, if sufficiently thick a photochemical haze layer could significantly cool a planet's surface and threaten habitability \citep{McKay1999, Haqq2008}.  Absorbing species in the atmosphere can also affect habitability by modifying the atmospheric thermal structure, which in turn can either help or hinder water loss via photolysis in the stratosphere \citep{Wordsworth2014, Fujii2017c}.  Planetary surface characteristics such as the presence of global oceans and the locations of continents, also can affect habitability through the modulation of ocean heat transports and their effect on the overall climate \citep{Hu2014, DelGenio2019, Yang2019}. 

Atmospheric properties will be probed via transit transmission spectroscopy, thermal phase curves, secondary eclipse and direct imaging spectroscopy \citep{Meadows2018}. 
Transit transmission spectroscopy can help us identify gas species in the upper atmosphere of planets \citep{Morley2017, Lincowski2018, Lustig-Yaeger2019}.  The detectability of a molecule depends on its atmospheric abundance, and the strength of spectral features, as well as the wavelength range observed, with some molecules more likely to be observed than others \citep{Schwieterman2015b}.  However, the presence of condensates, which generally present featureless spectra, may sharply obscure our ability to observe underlying gases with transmission spectroscopy \citep{Kreidberg2015, Lincowski2018,Morley2017,Lustig-Yaeger2019}.   Thermal emission and reflected light phase curves may yield clues to atmospheric composition, the presence of clouds and aerosols, and allow  temperature mapping \citep{Yang2014b, Koll2016, Wolf2019, Kreidberg2019}.  The temperature structure of atmospheres is more challenging to observe, but could be derived from thermal infrared spectroscopy that encompasses the 15 $\mu$m CO$_2$ band.  

\section{STELLAR CHARACTERISTICS FOR HABITABILITY}

The host star's characteristics have a huge influence  on a planet's environment and habitability.  Stellar mass and radius determine many of the star's fundamental characteristics, such as temperature and lifetime.  Stellar luminosity evolution drives strong climate change and may result in atmospheric or ocean loss, which is a compositional change and often a threat to habitability. The stellar spectrum and activity levels influence atmospheric escape and climate, provide the most abundant surface energy source for the majority of HZ planets, and photochemically modify the planet's atmospheric composition.  

\subsection{Luminosity, Age, and SED}

The energy emitted by the host star, and received by the planet, plays a primary role in determining whether a planet can be habitable. The luminosity of the star is a measure of the total energy it emits per unit time, and it depends on the star's size and emitting temperature.  The stellar luminosity controls the energy received by a planet, and in large part determines the semi-major axis of its HZ.

Stars have a finite lifetime, determined by their rate of fuel consumption. Smaller, cooler stars like M dwarfs have much lower luminosities than larger, hotter F dwarfs, and have nearly fully convective interiors that can deliver more fuel to the reacting core, so they burn at a low rate, but for longer. More massive stars support their higher luminosities by burning their atomic fuel at a much higher rate, but can't convect additional fuel to the core as efficiently as smaller stars, and so have significantly shorter life spans.  While our Sun, a G dwarf, may live for ~10  billion years, an A dwarf that is twice as massive as the Sun would remain on the main sequence for only ~2 billion years. This is significantly less time than it took for oxygen to rise to even 10\% of the current atmospheric level on our planet \citep{Lyons2014} and so produce a detectable biosphere.  M dwarf stars are small and dim, and can spend ~100s of billions of years on the main sequence (Rushby et al. 2013), far longer than the 13.8 Gy age of the Universe. 

Stars brighten over their main sequence lifetimes, thus the radiation received by an orbiting planet increases over time.  For instance, in Earth's early history it received ~$25\%$ less stellar energy than it does today.  Still, other factors controlling the atmospheric composition and the greenhouse effect allowed Earth to maintain continuously habitable surface temperatures despite this change in stellar irradiance over time (Robert and Chaussidon 2006).  Our Sun will continue brightening at a rate of ~$1\%$ every 100 million years (Gough 1981).  The rate of luminosity evolution depends on stellar mass, with larger hotter stars naturally brightening more rapidly than smaller, cooler stars.  


Lastly, the spectral energy distribution (SED) is the amount of radiation emitted by the star as a function of wavelength.  The SED is strongly dependent on the emitting temperature of the stellar photosphere.   The wavelength of peak stellar emission is inversely proportional to the emitting temperature.  G dwarfs, like our Sun, have peak emission at visible light wavelengths, while cooler stars, like M dwarfs, emit primarily near infrared radiation.  The relative SED of the star in turn can strongly influence the climate system, as surface reflectivity, near infrared gas absorption, and Rayleigh scattering are all sensitive to changes in the SED.  The relative SED may also strongly influence biology, as photosynthesis plays a dominate role in Earth's biosphere.

\paragraph{{Luminosity Evolution}}
 
Stars evolve in luminosity, from a super-luminous phase as they collapse down to their main sequence sizes, through gradual brightening as they fuse hydrogen into helium in their cores, increasing the core temperature and accelerating fusion.  For lower mass stars the pre-main sequence phase can be as long as 2.5 Gy, but is only 10My for a G dwarf like our Sun \citep{Baraffe2015}. This super-luminous pre-main sequence phase would subject planets that form in what will become the M dwarf's main sequence habitable zone to very large amounts of radiation early on, increasing the chance that these planets will experience ocean and atmospheric loss \citep{Luger2015b,Ramirez2014b,Tian&Ida2015}. For more Sun-like stars, the pre-main-sequence phase is relatively short, and the star’s luminosity then increases strongly during its main sequence phase (the Sun will undergo an 80\% increase in luminosity in its lifetime).  For the smaller M dwarf stars, the long  pre-main-sequence phase fades to an almost constant main-sequence luminosity over trillions of years.  While a star's current luminosity can be straightforwardly measured, its luminosity evolution is determined primarily using models that are validated against observed luminosities as a function of spectral type, mass and age \citep{Baraffe2015}.

\subsection{Activity Levels}

The stellar mass and age of the star will also affect the level of stellar activity, which can produce UV and shorter wavelength radiation that is potentially damaging to planetary atmospheres, ozone layers and surface life \citep{Wheatley2019,Tilley2017,Segura2010}. Stellar activity, including sunspots and flares, is produced by stellar magnetic field interactions, which are a function of the internal convection of the star and its rotation rate.   For solar-type stars a stellar magnetic field is generated via shearing due to differential rotation at the boundary between the radiative inner zone and the convective outer zone of the star's interior.  The field generated at this boundary rises buoyantly through the star's convective zone to emerge as magnetic loops on the stellar surface, which eventually release their magnetic energy in stellar flares.  For high-mass F dwarfs, the outer convective layer is too shallow for much field to be generated, but by M3V (stars $\leq 0.3$ M$_\odot$) stars become fully convective, and the magnetic field is generated instead by a turbulent dynamo, which can produce large-scale fields and strong stellar activity. The generation of the magnetic field is intimately linked to the stellar rotation, and this evolves as the star ages. Stars form with relatively high angular momentum and spin down over the course of their lifetimes so that young stars are more magnetically active than older stars, up until a characteristic  “saturation” spin velocity at which the observed activity appears to level out \citep{White2007,Gudel2004,West2009}. While early M dwarfs have solar-like spindown times and are inactive in just under a Gyr \citep{West2008}, fully-convective later-type M dwarfs have much longer spindown times, extending up to 8 Gyr for M7V stars \citep{Hawley2000,Gizis2002}.  Stellar activity levels and frequency are relatively easily measured using broadband photometry, but spectral information on the flares requires UV spectroscopy from space-based platforms such as GALEX or HST. There is a need for more EUV observations by future space missions to fully understand the impact of stellar activity on planetary habitability.

\section{PLANETARY SYSTEM CHARACTERISTICS FOR HABITABILITY}

In addition to the star-planet interaction, the planet may also interact with other components of the planetary system, during formation and subsequent evolution.  When assessing habitability, these planetary system components will need to be inventoried to better understand planet formation, volatile delivery and orbital modification. 

\subsection{Planetary System Architecture}

Other components of a planetary system, such as Jovian planets, asteroid and Kuiper belts, and nearby sibling planets can all impact the potential habitability of terrestrial planet, and provide clues to its formation and evolutionary history \citep{Raymond2008}.  The masses and orbits of Jovian planets in particular should be characterized, as they can affect volatile delivery to forming terrestrial planets.  Eccentric Jovians could result in the formation of water-poor terrestrials \citep{Raymond2004}, whereas Jovians that remain on wide orbits protect terrestrial planet formation in the inner planetary system while potentially enriching it with volatiles \citep{Raymond2017a}(see Chapters 12 and 13 in this book for an in depth discussion of the impacts of planet formation on terrestrial planet characteristics and habitability).  The presence of debris disks may indicate that an eccentric Jovian is not present \cite{Raymond2012}.  Nearby sibling planets can also modify orbital parameters including eccentricity and obliquity, and help maintain tidally-locked planets in 3:2 resonances, rather than the 1:1 resonance of synchronous rotation, although this may also occur if the planet has a very low ``triaxiality'' and is more truly spherical \citep{Ribas2016}.  Sibling perturbation from circular orbits and synchronous rotation may occur even for the closely packed systems seen orbiting M dwarfs, of which TRAPPIST-1 is a well known example \citep{Gillon2016,Gillon2017,Luger2017c}.  The 7 TRAPPIST-1 planets are found closely packed together in a resonant chain that implies migration from more distant birth orbits \citep{Luger2017c}, and their mutual gravitational interactions produce transit timing variations (TTVs) that have been used to determine their masses and densities, and constrain their orbital eccentricities to be less than 0.08 in most cases  \citep{Gillon2017}.  

Other planets in the planetary system can be observed in transit, via TTVs, with radial velocity, astrometry or direct imaging.   Belts of minor planets analagous to our asteroid and Kuiper belts serve as a reservoir for water-rich bodies and the disk's dust distribution can reveal collisions among these smaller bodies, or the gravitational signature of unseen planets.  These features of the planetary system may be detectable as infrared excesses or directly imaged \citep{Kraus2017}.  Exomoons can also influence planetary habitability by damping large obliquity oscillations for habitable worlds, but remain challenging to detect.  Future observations may look for changes in the center of the planet-exomoon composite image \citep{Agol2015Center}, or additional transit timing signals \citep{Kipping2011}

\section{STAR-PLANET-PLANETARY SYSTEM INTERACTIONS AND HABITABILITY}

The intrinsic properties of planets, stars, and planetary systems described above exert significant controls on planetary habitability.  However, the interactions among the planet, its host star, and its planetary system constitute another category of factors that in part determine whether a planet is and can remain habitable.  Radiative interactions with the host star can modify planetary atmospheric compositions by driving the photochemical production of aerosols or gas species.  These modifications of the atmosphere subsequently affect planetary climate and the UV flux incident at the planet's surface, both of which directly affect habitability.  In more extreme cases, oceans of water vapor and/or the atmosphere itself can be stripped away to space by EUV/XUV radiation and the stellar wind.  Gravitational interactions between the host star, planet, and system can modify orbital properties which in turn modulate insolation levels and therefore climate.  Gravitational interactions also may be responsible for late volatile deliveries from comets deflected into the inner part of stellar systems.  Tidal interactions between bodies in the system can influence planetary interiors, controlling the magnetic dynamo and plate tectonics, both of which play significant roles in the maintenance and retention of secondary atmospheres on terrestrial planets.  These processes will be better understood with an interdisciplinary systems approach to modeling terrestrial exoplanet environments.  Below we describe the significant planetary processes that are impacted by these interactions in more detail.

\subsection{Atmospheric and Ocean Loss and Replenishment}

Unlike the gas giants, none of the four terrestrial planets in our Solar System have retained their primordial H$_2$-dominated atmospheres. Instead they exhibit secondary atmospheres, composed of fractionated remnants of their primordial atmospheres augmented by outgassed volatiles from their interiors and volatiles delivered from comets and asteroids \citep{Pepin2006}. The loss of a primordial H$_2$ dominated atmosphere is probably beneficial for planetary habitability \citep{Luger2015a,Owen2016}.  For instance, the gas giants in our Solar system have retained their primoridal atmospheres, and as a consequence have dense and opaque atmospheres, immense pressures, and no true surfaces.  Any extant life on a gas giant would probably need to live among the clouds (e.g. \cite{Morowitz&Sagan1967}.  Still, the retention of a planet's secondary atmosphere remains of significant concern for the long term habitability of terrestrial planets.  In our own Solar system, we observe that Mars' secondary atmosphere has been stripped away over time \citep{Jakosky2018}, leaving insufficient atmospheric pressure remaining to keep Mars warm today despite it  being located within the habitable zone.  On the other hand, Venus has retained a thick secondary atmosphere of outgassed CO$_2$, but it has lost its water to space over time \citep{Kasting1983, Donahue1982}, rendering it water-poor and thus uninhabitable.

Atmospheric escape can be driven by several processes including XUV/EUV radiation from the star, stellar wind interactions, and erosion by impact events \citep{Ahrens1993,Quintana2016, Genda2005}.  Smaller planets, hotter planets, and planets without a protective magnetic fields will be more prone to atmospheric loss processes.  An analysis of the planets and moons in our Solar System, along with known exoplanets, suggests that bodies with and without atmospheres may be divided as a function of stellar insolation and escape velocity \citep{Zahnle2017}.  Large mass planets have stronger gravity and higher escape velocities, and thus are better able to resist atmospheric escape to space due to stellar EUV/XUV radiation and impacts.  Planets with magnetic fields are able to shield their atmospheres from stripping by the solar wind by deflecting charge particles around the planet.  

Planets around M dwarf stars may be particularly vulnerable to atmospheric loss via hydrodynamic escape processes due to their long ($>$Gy) super-luminous pre-main sequence phase and high activity levels \citep{Lammer2008,Luger2015b,Meadows2018,Barnes2018} as well as ion pickup \citep{Ribas2016}.  The strong stellar magnetic fields of M dwarfs also can reduce the size of planetary magnetospheres, exposing more of the planet's atmosphere to erosion by the stellar wind \citep{Vidotto2013}.  For magnetized planets orbiting M dwarfs, the calculated polar wind losses for a 1-bar Earth-like atmosphere suggested a lifetime for that atmosphere of less than 400 Myr, although the calculated loss rate of H$_2^+$ and O$_2^+$ did not exceed Earth’s current replenishment rate via outgassing and volatile delivery, so that maintenance of the atmosphere might be possible \citep{Garcia2017}. However, planets without a magnetic field will be more vulnerable to atmospheric loss, as may be the case for older synchronously-rotating M dwarf planets in the habitable zone \citep{Driscoll2015}.  If enough atmosphere is lost via interaction with the star, then the entire atmosphere could potentially condense onto the cold nightside \citep{Joshi1997, Leconte2013b, Turbet2017} or at the poles of the planet \citep{Turbet2017b}.  Planets that have completely lost their atmospheres could be identifiable by the presence of extreme day-night temperature differences \citep{Kreidberg2019}.
 
Venus, on the other hand, poses the scenario where many oceans of water can be lost to space over time, even when other atmospheric constituents remain.  D/H ratios suggest that Venus has lost significant amounts of water to space over its lifetime \citep{Donahue1982, Donahue1999}, while climate modeling studies suggest that Venus could have had clement surface conditions during the early history of our Solar System when the Sun was dimmer \citep[][and see chapter by Arney and Kane in this volume]{Yang2014b, Way2016}.  Independent of the bulk stripping of the atmosphere, water loss to space for Venus and other planets occurs when the climate enters a moist or runaway greenhouse state \citep{Kasting1988}, both of which allow large quantities of water vapor to permeate a planet's stratosphere where it can be photolyzed and the freed H atoms then irreversibly escape to space.  H atoms, being the lightest of all atoms, can escape to space most easily while heavier molecular gases like N$_2$ and CO$_2$ remain bound to the planet.  Habitable planets around M dwarf stars may be at greater risk of water loss during moist greenhouse states due to their particular patterns of atmospheric circulation \citep{Kopparapu2017, Fujii2017}; however, ultimately water loss rates may depend on the level of stellar activitiy \citep{Chen2019}.

The loss of oceans to space could significantly modify the composition of a planet's atmosphere \citep{Luger2015b}.  Photolysis of H$_2$O and the subsequent removal of H results in the build up of approximately 250 bars of O$_2$ for each Earth ocean equivalent of water lost. The amount of O$_2$ that remains in an atmosphere over long time scales depends on further loss processes, including atmospheric loss, and losses to sequestration in a magma ocean or via other {surface processes} \citep{Schaefer2016, Wordsworth2018}.  This potential build up of abiotic O$_2$ is a potential false positive biosignature \citep{Harman2015}.

While ocean loss process are themselves irreversible, the effects of ocean loss during moist and runaway greenhouse climate states may be mitigated by planets forming with higher initial water abundances \citep{Tian&Ida2015}, planetary migration from beyond the snowline \citep{Luger2017c}, the presence of a dense protective H$_2$ atmospheres early on \citep{Luger2015a,Barnes2018}, and subsequent cometary delivery or outgassing \citep{Albarede2009, Meadows2018}.  For example, despite the Earth having lost 80-95\% of its volatiles within the first 50-500Myr \citep{Turner1989}, large quantities of water may have remained sequestered in the mantle and were then slowly outgassed over time \citep{Sleep2012c}.  Volatile cycling between the mantle and atmosphere may take billions of years to reach a steady-state, allowing exoplanets to regain surface volatiles that were lost during the early phases in their history \citep{Komacek2016}. If M dwarf terrestrial exoplanets can similarly acquire volatiles after an initial loss of water and atmosphere, they may, over billions of years, accumulate a surface ocean and atmosphere from volcanic outgassing or volatile delivery, after the M dwarf has settled into its more benign main sequence phase.

\subsection{Photochemistry}

The composition of terrestrial exoplanet atmospheres can be strongly affected by photochemistry \citep{Segura2005,Rugheimer2015, Meadows2018, Lincowski2018}.  Photochemistry refers to gas-phase chemistry that occurs in a planet's atmosphere and is driven by light of the host star.  Photochemistry is driven by the stellar UV spectrum, and the resultant gases and aerosols depend on the initial atmospheric composition and temperature structure, the total amount of UV emitted by the star, and the wavelength dependence of the star's UV output \citep{Rugheimer2015}.  Upper-atmospheric chemistry can also be driven by stellar energetic particle precipitation from coronal mass ejections \citep{Airapetian2016}.  Some molecules can be directly photolyzed, as is the case for CH$_4$, N$_2$O, and H$_2$O, while the abundance of other molecules (including CH$_4$) can be highly sensitive to the presence of catalysts.  The concentration of catalysts may be independently controlled by both photochemical \citep{Segura2005} and dynamical processes \citep{Schoeberl1991}.  

Cooler main sequence stars often emit less UV and thus are often not as efficient as the Sun at photolyzing water vapor.  This has a variety of consequences including, the reduced production of the O($^1$D) catalyst that is efficient at chemically destroying CH$_4$ \citep{Segura2005}, and the reduced production of OH radicals required for CO$_2$ recombination \citep{Harman2015}.  Planets around quiet M dwarf stars may experience only limited water loss from moist greenhouse climate states due to inefficient photolysis, however water loss rates may be critically dependenty on the level of stellar activity \citep{Chen2019}.  Photochemically produced O$_3$ in oxygen-rich atmospheres and Titan-like hydrocarbon hazes in methane-rich atmospheres can both provide {UV shielding} of the planetary surface by absorbing UV radiation high in the atmosphere \citep{Arney2016}.  However, note that some photochemical modifications to the atmosphere can be harmful.  For instance, planets around M dwarfs stars might build up significant abundances of CO in their atmospheres, which is poisonous gas to complex life as we know it \citep{Schwieterman2019}.

\subsection{Climate}

As discussed above, if water is to remain liquid at the surface, a planet must maintain a climate state that supports surface temperatures that lie in a fairly narrow range.  A planet's climate is determined through a delicate balance between absorbed incoming stellar radiation and emitted thermal radiation, modulated by the presence of greenhouse gases, clouds, and aerosols \citep{Trenberth2009}.  However, star, planet, and planetary system interactions can have important influences on both patterns of stellar insolation and on the composition of planetary atmospheres.  

Gravitational interactions can also drive changes to planetary climate via orbital modifications including migration and synchronous rotation.  Long term gravitational interactions can drive planet migration which changes the mean insolation received by a planet.  Giant planet migration may also be responsible for disturbing the orbits of comets and asteroids, triggering a bombarbment and delivery of fresh volatiles to the inner terrerstrial planets, shaping the composition of their secondary atmospheres \citep{Mojzsis2019}.  Gravitational interactions among neighboring bodies can drive oscillations in planetary eccentricity and obliquity \citep{Spiegel2010,Armstrong2014,Brasser2014,Deitrick2018a} which can affect the seasonal variability of planetary surface temperatures \citep{Bolmont2016, Rose2017, Kilic2017, WayGeorgakarakos2017, Adams2019, Colose2019}.  While the time-mean climates of highly eccentric planets may be stable \citep{Bolmont2016}, dramatic seasonal swings in surface temperature or ice cover could make sustained surface habitability challenging \citep{SherwoodHuber2010}.  Furthermore, planets in the habitable zones of M dwarf stars likely experience tidal locking where the planet maintains a synchronous or resonant orbit, thus governing the planet's rotation rate, and significantly changing atmospheric circulation patterns and climate \citep{Yang2014b}.

Photochemistry, driven by the stellar energy distribution and stellar activity level of the host star, can cause for the chemical modification of terrestrial planet atmospheres \citep{Segura2010, Arney2017}, which will impact temperature structure, climate and water loss.   Photochemistry impacts planetary climate, and thus habitability, by destroying or creating greenhouse gases, by creating absorbing aerosols species, and by driving planetary water loss process by removing H from high altitude water vapor (thus permitting H to escape to space).  The photolysis and ultimate destruction of water vapor in terrestrial planet atmospheres can yield a dessicated planet which has very different climate dynamics compared to water-rich planets \citep{Abe2011, Leconte2013b, Kodama2015}.  Photochemistry can also drive the creation or destruction of different novel greenhouse gases species in a planetary atmosphere such as CH$_4$, C$_2$H$_6$, NH$_3$ or N$_2$O which all can act as greenhouse gases and cause warming of the planet \cite{Segura2005, Haqq2008, Airapetian2016, Meadows2018}.  Conversely, photochemically produced upper atmospheric hazes can have a potentially significant cooling effect on climate \citep{McKay1999, Haqq2008, Arney2016}.  The addition of greenhouse gases or atmospheric hazes can either help or hinder planetary habitability depending upon the initial climate state.

\subsection{Tidal Effects}

Planets close to their host stars are expected to be affected by the differential gravitational force, and experience tides which can affect a planet's habitability \citep{Rasio1996,Jackson2008}. In particular, HZ planets around low-mass stars, such as M-dwarfs, may experience tidal forces which can deform their solid bodies, in addition causing changes in angular momenta (rotation) and energy that affect their atmospheric dynamics (Yang et al. 2013a; Kopparapu et al. 2016; Wolf 2017; Haqq-Misra et al. 2018). 

In addition to tidal locking \citep{Dole1964,Barnes2017}, which can result in synchronous rotation, tides may also impact habitability via orbital circularization, orbital migration, obliquity erosion, and tidal heating. 
Tides will also drive planetary obliquities towards 0 or 180° \citep{Goldreich1966,Heller2011}, which  changes the insolation pattern and may lead to atmospheric collapse at the poles \citep{Joshi1997}. Near the end state of tidal evolution, planets may become {tidally-locked}, and once their eccentricity and obliquity are both near zero they may synchronously rotate with one side of the planet always facing the star.  However, close-in planets can also be found in non-circular orbits, because either they have not yet been tidally circularized, or they have had their eccentricity or obliquity maintained over long time periods via perturbation by another planet---which  counteracts the tendency of tides to damp eccentricity and obliquity to zero \citep{Barnes2010}.  Planets in non-circular orbits can be heated by friction induced by the changing deformation of the planet \citep{Jackson2008,Barnes2009a}. Planets in the habitable zones of M dwarfs may experience Io-like levels of surface heat flux, which could significantly change their internal properties and outgassing rates, and potentially induce a runaway greenhouse \citep{Barnes2013a}.  Tidal heating can also induce a prolonged magma ocean stage for young, close-in planets before they circularize \citep{Driscoll2015}, and dense atmospheres may prolong this magma ocean phase \citep{Hamano2013}. This tidal heating of the mantle can promote core cooling and generation of an early magnetic dynamo, which is maintained as the planet circularizes and loses its tidal heating component.  However, if eccentricity is maintained for prolonged periods then the mantle cannot cool, which results in core solidification and loss of the magnetic dynamo, exposing the planet to stellar wind erosion. For these hotter eccentric planets, massive melt eruptions may also render them uninhabitable \citep{Driscoll2015}.

\subsection{Galactic Effects}

The planetary system that a habitable planet forms in also interacts with the Galaxy, and these interactions may affect the composition of the host star and protoplanetary disk, the dynamical stability of the planetary system, and the radiation received by the planet (for a more detailed review of Galactic impacts on habitability, see \citet{Kaib2018}).  Initial observations of a correlation between stars with enhanced metallicity (elements heavier than hydrogen) and the formation of giant planets \citep{FischerValenti2005} implied that perhaps planets would be more frequent in the inner, higher metallicity regions of the Galaxy.  However, subsequent studies showed no such correlation for planets with R $< 4$ R$_\oplus$ \citep{Buchhave2012, BuchhaveLatham2015}, which are found around stars with a range of metallicities that are much wider than expected, based on the hot Jupiter results.  Consequently there is no clear constraint on the location of terrestrial planet formation in our Galaxy as a function of stellar metallicity, and by extension, the metallicity of the interstellar medium.  

Galactic metallicity may also be a less important factor in supporting a planet with active plate tectonics, which is part of the carbonate-silicate cycle that buffers planetary climate \citep{Walker1981}.  Planet formation in a region of the galaxy that does not have recent star formation and regular type-II supernovae, which generate the $^{235,238}$U, $^{232}$Th and $^{40}$K unstable isotopes thought to drive radiogenic internal energy for tectonics, was also thought to limit the formation of terrestrial planets with sufficient long-lived energy for plate tectonics.  However, neutron star mergers, which are less tied to very recent star formation, and for which gravitational wave signatures were recently detected \citep{Abbott2017}, are now recognized to be a dominant contributor to the production of a more constant supply of unstable isotopes of U and Th. In addition, there is still uncertainty as to the relative contributions to Earth's current internal heat, and radiogenic decay may produce a heat flux that is roughly equal to the Earth's primordial heat of formation \citep{Korenaga2008,Dye2012}, such that even a complete lack of heating from radionuclides may  not significantly drop the surface heat flow.    

Gravitational interactions with the Galactic environment can perturb planetary systems, and possibly contribute to significant {migration} of the host star from its birth environment, ultimately affecting habitability.  While still in their formation clusters, stars are closer to each other and have lower relative velocities than they will out in the field, and so gravitational encounters between stars are more likely to perturb planetary orbits, including ejecting planets \citep{Laughlin1998,Spurzem2009}. However, this is unlikely to be an issue for most habitable zone planets, as an extremely close pass by another star would be needed to eject an inner planet, and most star-forming clusters disperse on 10 My timescale, which lowers the probability of such an encounter.  Although it would be challenging to eject a habitable zone planet, outer planets are more likely to be perturbed and the modification to their orbits could be transferred to inner planets via planet-planet interactions (Malmberg et al., 2011).  After the open cluster phase, stellar interactions would normally be less frequent, but if the host star has a distant stellar companion, that companion can act as an antenna for gravitational encounters with other stars.  This can ultimately destabilize the planetary system, and make habitability for planets in widely space binaries less likely over long time periods \citep{Kaib2013}.  On an even larger scale, stars on nearly circular orbits co-rotating with the Galaxy's spiral structure may be susceptible to large radial migrations \citep{Sellwood2002,Roskar2012}. Sun-like stars could have migrated up to 6 kiloparsecs (kpc) within the galactic disk, from the inner Galaxy outward, significantly changing the radiation environment over time, and the stellar encounter frequency \citep{Wielen1996,Roskar2011}.  Increased stellar encounters when the Sun was younger and closer to the center of the Galaxy could have influenced the volatile delivery and impact history of Earth by modifying the Oort Cloud and the flux of comets through the inner Solar System \citep{Kaib2011}.  These interactions could have reduced or destroyed reservoirs of distant icy bodies or  injected outer planetary system icy bodies into habitable orbits. 

Finally, the galactic environment can affect the planet directly, via interaction with radiation and particles from highly energetic events. Proposed mechanisms include the action of gamma ray bursts \citep{Melott2004,Atri2014}, supernovae \citep{Gehrels2003} or kilonovae \citep{Abbott2017} which could erode the ozone layer on a habitable planet. However, models of the effect of a nearby (8pc distant) supernova suggest that surface UV flux would only increase by a factor of 2, which may not be catastrophic.  Moreover, such a nearby supernova would likely only happen once every 8 Gy \citep{Gehrels2003}.  Gamma ray bursts are far more energetic, and could indeed produce mass extinction events if a habitable planet were exposed to one within 1-2 kpc \citep{Thomas2005}. However, GRBs are most often observed in metal poor (less than 10\% of the Sun's metallicity)  galaxies \citep{PiranJimene2014}, and if their progenitors are similarly metal poor, very few stars would have been subjected to these GRBs, and those would be in the more metal poor outskirts of the Galaxy \citep{Gowanlock2016}. Consequently, the impact of gamma ray bursts and supernovae on planetary habitability may be relatively modest, especially when compared to processes that are likely to have a higher impact, e.g. stellar activity, within a planetary system.  

\section{Current and Near-Term Observations of Habitable Zone Planets}\label{sec:observations} 

We are now entering a new era of terrestrial exoplanet characterization that is providing a glimpse into the environments of possibly habitable worlds.  Although models may provide valuable information about the probability of habitability, it is observations, albeit often interpreted by models, that will ultimately be used to assess whether or not a planet is habitable.  In the past few years, the very first attempts at determining planetary bulk composition, and observing the atmospheres of likely terrestrial exoplanets in the habitable zones of M dwarfs have been undertaken.  A handful of good targets are known, including the HZ transiting planets TRAPPIST-1 e, f and g in the 7-planet TRAPPIST-1 system \citep{Gillon2016,Gillon2017} and LHS 1140 b \citep{Dittmann2017} and the RV-detected Proxima Centauri b \citep{Anglada-Escude2016}, and Ross 128 b \citep{Bonfils2018}.  To complement these HZ planets, other likely terrestrials include the exo-Venuses TRAPPIST-1 b, c and d \citep{Gillon2016,Gillon2017} and GJ1132 b \citep{Berta2015}.  For the transiting planets, observations of both the size and inclination from transit, combined with masses from RV or TTVs have produced densities, that provide our first steps towards characterizing habitability. The densities of nearby M dwarf terrestrial planets (e.g., GJ1132 b, $6.0\pm 2.5$ g  cm$^{-3}$, \citep{Berta2015}; and LHS1140 b, $12.5\pm 3.4$ g cm$^{-3}$, \citep{Dittmann2017} are comparable to the densities of Earth (5.5 g  cm$^{-3}$) or Venus (5.3 g cm$^{-3}$), consistent with mixtures of silicate rock and iron.  Initial estimates suggested that TRAPPIST-1 planets have densities that span 0.6 to 1.0 times Earth’s density (i.e., $3.3-5.5$ g  cm$^{-3}$) \citep{Grimm2018}, and innovative techniques that used data from the seven planets together to probe their interiors suggested that the measurement errors were constant with a consistent or increasing water mass fraction with semi-major axis \citep{Dorn2018}. Although new measurements with better precision suggest that the densities of most of the TRAPPIST-1 planets are in fact similar to each other, and are closer in value, albeit with higher water fractions, to our Solar System terrestrials. The generally lower densities of the TRAPPIST-1 planets \citep{Gillon2017,Grimm2018}, along with their resonant orbits, suggest that the planets have formed at larger distances from the star in a more volatile-rich birth environment and migrated inward   \citep{Luger2017b, Unterborn2018}.  So despite predictions of extreme atmospheric and ocean loss for these planets, it is likely that their interiors remain volatile rich. 

Spectra and photometry have also been obtained for several of the TRAPPIST-1 planets, and although these spectra appear to be featureless, they do help in attempts to constrain the planetary atmospheric properties. HST Wide Field Camera 3 transmission spectroscopy has ruled out H2-dominated atmospheres for the innermost six TRAPPIST-1 planets \citep{DeWitt2016, DeWitt2018}. Additionally, laboratory data and models suggest that it is unlikely that the flat spectra observed are due to suspended aerosols, and instead may be high mean-molecular-weight secondary outgassed atmospheres \citep{Moran2018}, of as yet unconstrained composition \citep{Delrez2018, Lincowski2018}.  However, whether the planets have high molecular weight atmospheres (e.g. CO$_{2}$- or O$_{2}$-dominated) or no atmospheres requires observations with future facilities

In the next 5-10 years, JWST, which is scheduled for launch early in 2021, and next-generation ground-based extremely large telescopes will provide additional capabilities to study the atmospheres of HZ terrestrial exoplanets.  The best targets for JWST will likely come from ground-based exoplanet detection surveys that focus on late-type M dwarfs, as these small stars produce excellent atmospheric signals for transiting planets \citep{Morley2017,Lustig-Yaeger2019}.  Planets discovered by the TESS mission are more likely to be orbiting brighter, earlier type M dwarfs that produce relativley weak differential signals from planetary atmospheres and so will not be the best targets for JWST.  However, for systems like TRAPPIST-1, which is orbiting a late-type M8 dwarf, JWST can likely detect the presence of a terrestrial atmosphere (containing CO$_2$ but no aerosols) for each of the planets using transmission spectroscopy, by coadding data from fewer than 10 transits for each planet \citep{Morley2017,Lustig-Yaeger2019}.  For cloudy atmospheres the integration time will be up to 20-30 transits \citep{Lustig-Yaeger2019}.  For characterizing the nature of the atmosphere, and the a planet's habitability, JWST will be much more adept at proving ocean loss, or a loss of habitability, than confirming that an ocean is present.  This is because signs of ocean loss, such as strong O$_{2}$-O$_{2}$ collisional absorption from a massive O$_{2}$ atmosphere \citep{Meadows2017a,Lincowski2018}, as well as potential signatures from enhanced D/H from atmospheric loss are potentially  detectable in 2-11 transits for the inner planets.  The presence of gases that are normally soluble in water, such as SO$_2$, or the presence of an  H$_{2}$SO$_{4}$-H$_{2}$O haze layer with SO$_{2}$ gas \citep{Lincowski2018,Loftus2019} are more observationally challenging \citep{Lustig-Yaeger2019,Loftus2019} but would point to a surface environment almost entirely devoid of water.  On the other hand, water vapor, at the relatively desiccated altitudes that transit probes, will take upwards of 60 transits for the HZ TRAPPIST-1 planets \citep{Lustig-Yaeger2019,Lincowski2019}, and even if detected it is not definitive proof that surface liquid water exists.   Biosignatures like O$_{2}$ will be extremely challenging for JWST and unlikely to be detected due to  the poor sensitivity of the instruments at wavelengths shortward of 1.3$\mu$m.   

Ground-based extremely large telescopes (ELTs), may also be capable of probing M dwarf planetary atmospheres, starting in ~2025. The ELTs can probe a handful of HZ planets using high-resolution spectroscopy for transiting \citep{RodlerLopez-Morales2014} and reflected light observations \citep{Snellen2015, Lovis2017}.  These telescopes may also be able to take direct imaging mid infra-red (MIR) observations of planets orbiting G dwarf stars \citep{Quanz2015}. For the best targets, these facilities may have the precision required to undertake the first spectroscopic search for atmospheric water vapor and biosignature gases, such as O$_2$ and CH$_4$ \citep{Lovis2017, RodlerLopez-Morales2014,Lopez-Morales2019}, but these ground-based measurements will also be unable to directly detect surface water on an exoplanet.  

In the more distant future, large space-based coronagraphic direct imaging mission concepts currently under consideration by NASA may obtain spectra of 100s of habitable zone planets. These planets will be orbiting stars of spectral type from F down through M and will provide a large statistical sample for observational determination of the habitable zone, and detection of biosignature gases. Constraints on CO$_{2}$ levels on the larger samples of planets obtained by these telescopes could be used to test for evidence of the carbonate-silicate cycle directly \citep{Bean2017}. These telescopes may also have the capability to map the planetary surface as it rotates under the observer, and directly detect glint from exoplanet oceans \citep{Robinson2010, Cowan2012, Lustig2018}, providing a definitive detection of habitability. For a more detailed discussion of how to observationally determine exoplanet habitability, see Chapter by Robinson \& Reinhard in this volume.

\section{SUMMARY AND CONCLUSIONS}

The characteristics and processes relevant to the maintenance of surface liquid water on a terrestrial planet are broad, interdisciplinary and interconnected, and both modeling and observations will be needed to understand them.  To date, our best first-order assessment method for whether or not a planet is likely to be habitable has been to check whether a newly discovered exoplanet is in the size range that is likely to be terrestrial, and is in the habitable zone of its parent star.  However, we now know that habitability is maintained via the interplay of planetary, stellar and planetary system characteristics over the planet's lifetime. Within this new framework, the habitable zone can be seen as a 2-dimensional slice in stellar type and semi-major axis through a multi-dimensional parameter space.  Planets that form in the HZ may also not be habitable due to either a low initial volatile inventory, or loss of volatiles over time.  Atmosphere and ocean loss is of particular concern for M dwarf terrestrial planets which orbit close to a star that spent its early life in a super-luminous state and  maintains high stellar activity for an extended period of time.  On the other hand, terrestrial planetary atmospheres are most likely secondary, outgassed from the interior.  Understanding how the balance between outgassing and atmospheric escape sculpts the resulting terrestrial planet atmosphere, and potentially replenishes an ocean,  will be an important new frontier in terrestrial exoplanet evolution and habitability.  While some of the characteristics and processes that inform planetary habitability may be observable in the coming decades, many will instead be explored via modeling, or a combination of modeling and observations.  An interdisciplinary system science approach will be needed to fully explore the depth and complexity of planetary habitability.  An improved understanding the factors that affect habitability will enable identification of those exoplanets that are most likely to be habitable, and inform our interpretation of upcoming exoplanet data to be used to search for life beyond the Earth.

\section{Acknowledgements} We sincerely thank reviewers Dorian Abbot and Robin Wordsworth for their constructive comments and suggestions, which greatly improved the review chapter. This chapter benefited from discussions with Andrew Lincowski, and the assistance of Evan Davis. This work was performed by NASA's Virtual Planetary Laboratory, a member of the Nexus for Exoplanet System Science Research Coordination Network, and  funded by the NASA Astrobiology Program under grant 80NSSC18K0829.
\parskip=0pt
{\small
  \baselineskip=11pt
  \bibliographystyle{apalike}
\bibliography{factors}

}

\end{document}